\documentclass[preprint]{aastex}

\usepackage{graphicx,amsmath,amssymb}

\begin{document}
\title{Velocity-Resolved [Ne {\sc iii}] from X-Ray Irradiated Sz 102 Microjets}
\author{Chun-Fan Liu\altaffilmark{1,2,3}, Hsien Shang\altaffilmark{1,3},
        Frederick M. Walter\altaffilmark{4}, and Gregory J. Herczeg\altaffilmark{5}}
\altaffiltext{1}{Institute of Astronomy and Astrophysics, Academia Sinica (ASIAA), 
P. O. Box 23-141, Taipei 10641, Taiwan}
\altaffiltext{2}{Graduate Institute of Astrophysics, 
National Taiwan University, Taipei 10617, Taiwan}
\altaffiltext{3}{Theoretical Institute for Advanced Research in Astrophysics
(TIARA), Academia Sinica, P. O. Box 23-141, Taipei 10641, Taiwan}
\altaffiltext{4}{Department of Physics and Astronomy, Stony Brook University, 
Stony Brook, NY 11794-3800, USA}
\altaffiltext{5}{The Kavli Institute for Astronomy and Astrophysics, 
Peking University, Beijing 100871, China}

\begin{abstract}
Neon emission lines are good indicators of high-excitation regions close to a young stellar system because of their high ionization potentials and large critical densities. We have discovered [Ne {\sc iii}] $\lambda3869$ emission from the microjets of Sz 102, a low-mass young star in Lupus III. Spectroastrometric analyses of two-dimensional [Ne {\sc iii}] spectra obtained from archival high-dispersion ($R\approx 33,000$) Very Large Telescope/{\sc Uves} data suggest that the emission consists of two velocity components spatially separated by $\sim0\farcs3$, or a projected distance of $\sim60$ AU. The stronger redshifted component is centered at $\sim +21$ km\,s$^{-1}$ with a line width of $\sim 140$ km\,s$^{-1}$, and the weaker blueshifted component at $\sim -90$ km\,s$^{-1}$ with a line width of $\sim 190$ km\,s$^{-1}$. The two components trace velocity centroids of the known microjets and show large line widths that extend across the systemic velocity, suggesting their potential origins in wide-angle winds that may eventually collimate into jets. Optical line ratios indicate that the microjets are hot ($T\lesssim1.6\times10^4$ K) and ionized ($n_e\gtrsim5.7\times10^4$ cm$^{-3}$). The blueshifted component has $\sim13\%$ higher temperature and $\sim46\%$ higher electron density than the redshifted counterpart, forming a system of asymmetric pair of jets. The detection of the [Ne {\sc iii}]$\lambda3869$ line with the distinct velocity profile suggests that the emission originates in flows that may have been strongly ionized by deeply embedded hard X-ray sources, most likely generated by magnetic processes. The discovery of [Ne {\sc iii}] $\lambda3869$ emission along with other optical forbidden lines from Sz 102 support the picture of wide-angle winds surrounding magnetic loops in the close vicinity of the young star. Future high sensitivity X-ray imaging and high angular-resolution optical spectroscopy may help confirm the picture proposed.
\end{abstract}


\section{INTRODUCTION} \label{introduction}

       Understanding the launching processes and the excitation mechanisms is crucial in studies of jets and outflows from low-mass young stellar objects (YSOs). Theoretical studies have shown that jets are launched from inner disks of young stars. High-resolution spectroscopy and imaging have revealed forbidden emissions from ``microjets'' on scales of a few tens of AU. High-dispersion spectroscopy has helped to examine these properties through velocity-resolved line profiles and line ratio diagnostics even with moderate spatial resolutions. 
       
       The low-mass YSO Sz 102 \citep{Sz77}, located in the Lupus III cloud, was first identified as an emission-line star (Th 28) by \citet{The62}. Its $377$ \AA\ equivalent width H$\alpha$ emission line is among the largest in the Lupus star-forming regions \citep{Hughes94}. Low-resolution optical spectroscopy of Sz 102 shows forbidden emission lines resembling those from Herbig--Haro (HH) objects, superimposed on a weak red continuum \citep{Krautter84}. The stellar emission from Sz 102 is underluminous ($0.03\,L_\odot$) for a K- or M-type pre-main sequence star \citep{MOD11}, which suggests that the disk is observed close to edge-on and obscures the star. The observed spectral energy distribution, from which a disk thermal emission of $\sim0.13\,L_\odot$ was derived \citep{MOD11,Merin08}, has a rising infrared index \citep[$\alpha\approx0.72$;][]{Chapman07,Evans09}.

       The geometry inferred from outflow knots is also consistent with that inferred from the continuum emission. Along the position angle of 98\degr, at least three HH knots were identified, two on the east side and one on the west side, designated as HH 228 E1, E2, and W \citep{Krautter86,GH88}. Spectroscopy of [S {\sc ii}] and [N {\sc ii}] emission lines shows that knots E1 and E2 are blueshifted from the systemic velocity by $-67$ and $-87$ km\,s$^{-1}$, and knot W is redshifted by $+33$ km\,s$^{-1}$ \citep{GH88}. Proper motions of the HH knots are $\sim0\farcs5$\,yr$^{-1}$ for the blueshifted knots and $\sim0\farcs4$\,yr$^{-1}$ for the redshifted knots. With an inferred distance ranging between 150 and 200 pc, the outflow is inclined at $\sim 5\degr$ -- $10\degr$ from the plane of sky \citep{Krautter86,WH09,CF10}. This corresponds to an inclination angle between $80\degr$ and $85\degr$.
      
    Spectra of Sz 102 show signatures of asymmetry in its pair of microjets. The east and west components have projected lengths of $\sim12\farcs4$ and $\sim13\farcs8$ \citep{Krautter86} and radial velocities $\sim-78$ and $\sim+23$ km\,s$^{-1}$ relative to the star \citep{GH88}, respectively. High-dispersion ($R\approx33,000$) spatially unresolved spectra with the {\sc Ultraviolet and Visual Echelle Spectrograph (Uves)} on the Very Large Telescope (VLT) by \citet{CF10} show most of the forbidden line profiles consist of a strong peak at $\sim+23$ km\,s$^{-1}$ and a blueshifted wing extended to $\sim-350$ km\,s$^{-1}$. \citet{CF10} interpreted the redshifted peak as the receding microjet but the high-velocity blueshifted wing as an uncollimated stellar wind. 

      The physical conditions of the Sz 102 microjets have been studied with optical spectroscopy at various spectral resolutions. Low-dispersion spectra show strong nebular lines such as [O {\sc i}] $\lambda\lambda6300,6363$, [N {\sc ii}] $\lambda\lambda6548,6583$, and [S {\sc ii}] $\lambda\lambda6716,6731$. High-excitation lines including [O {\sc iii}] $\lambda\lambda4959,5007$ and [S {\sc iii}] $\lambda6312$ are also detected. High electron density $\gtrsim 10^4$ cm$^{-3}$ is indicated by strong [S {\sc ii}] $\lambda\lambda4068,4076$ and moderate [O {\sc i}] $\lambda5577$ and [N {\sc ii}] $\lambda5755$ lines \citep{Krautter84}. The redshifted component was studied by \citet{BE99} at a spatial scale of $\sim2\arcsec$ and by \citet{Cof08} at $\sim0\farcs3$ from the source, while the blueshifted component is too faint for a reliable inference. The [S {\sc ii}] $\lambda\lambda6716,6731$ doublet ratio is saturated close to the base of the red component, suggesting a high electron density $\gtrsim 1\times10^4$ cm$^{-3}$. The average temperature inferred from [O {\sc i}] $\lambda6300$/[S {\sc ii}] $\lambda6731$ reaches $2\times10^4$ K and ionization fraction inferred from [N {\sc ii}] $\lambda6583$/[O {\sc i}] $\lambda6300$ gives a high value of $\sim0.3$ at $0\farcs3$ \citep{Cof08}. 
      
      Here we reanalyze the archival VLT/{\sc Uves} spectra of Sz 102, using the spatial sampling along the slit to separate the velocity components and examine the spectrally-resolved kinematics. In particular, we focus on the [Ne {\sc iii}] $\lambda$3869 emission.

      A noble gas, neon has high ionization potentials, so its emission lines of multiple ionization states are good indicators of high-energy processes. Ionization of the first valence electron requires 21.6 eV and the second for 41.0 eV. Alternatively, neon can be ionized through inner-shell photoionization, which requires 0.87 and 0.88 keV for $K$-shell electron ejection for neutral and singly-ionized neon, respectively. These properties suggest that photons within or beyond the energy range of the extreme ultraviolet (EUV, $13.6$ eV $<h\nu \lesssim100$ eV) are necessary for ionizing the outer electrons \citep{GH08}. Hard X-rays with energies on the order of keV are important and efficient in neon ionization since the $K$-shell photoionization has its maximal cross section at $\sim0.9$ keV and it is followed by consequent Auger electron ejections \citep{GNI07}. Jump shocks stronger than 100 km\,s$^{-1}$ \citep{HG09}, which typically occur at large bow shocks in HH objects \citep{HRH87}, provide another possible source of ionization. 
 
       Sz 102 is one of the six YSOs for which both [Ne {\sc ii}] 12.81\micron\ and [Ne {\sc iii}] 15.55\micron\ were detected in the {\it Spitzer}/Infrared Spectrograph (IRS) surveys \citep{Lahuis07,Esp13}. Although the low mid-infrared (MIR) [Ne {\sc iii}]/[Ne {\sc ii}] ratio \citep[0.064,][]{Lahuis07} is consistent with model predictions of an X-ray--irradiated disk and photoevaporative wind \citep{MGN08,EO10}, other evidence suggests that neon lines from Sz 102 may be associated with its microjets \citep{Shang10}. \citet{Gudel10} demonstrated a clear bimodal distribution of [Ne {\sc ii}] 12.81\micron\ luminosities among those YSOs detected with the IRS. Along with other jet-driving sources including T Tau and DG Tau, Sz 102 has a larger [Ne {\sc ii}] 12.81\micron\ luminosity that is roughly 1 dex larger than the average value of $\sim 2\times10^{28}$ erg\,s$^{-1}$ from disk sources. Calculations with X-wind jet models in \citet{Shang10} support the notion that the higher luminosities of [Ne {\sc ii}] 12.81\micron\ line in the bimodal distribution correlate well with the presence of stronger [O {\sc i}] $\lambda6300$ emission, which is an indicator of jet activity. Additionally, [Ne {\sc ii}] emission was not detected in Sz 102 with the VLT/VISIR long-slit spectroscopy, as the slit orientation was perpendicular to the jet axis. This non-detection supports the interpretation that the IRS-detected [Ne {\sc ii}] 12.81\micron\ might come from the jet \citep{PS09}. Further constraints require velocity-resolved spectra of both [Ne {\sc ii}] and [Ne {\sc iii}] in the MIR. Since high-dispersion observations of the MIR [Ne {\sc iii}] line are not viable from the ground, the optical [Ne {\sc iii}] $\lambda\lambda3869,3968$ doublet, with critical electron density of $\sim10^7$ cm$^{-3}$, stands out as a critical tracer for the jet origins of neon emission from low-mass YSOs.
      
        In this work, we report the velocity-resolved [Ne {\sc iii}] $\lambda3869$ line from the Sz 102 microjets. This is the first detection of optical [Ne {\sc iii}] $\lambda3869$ emission in the near vicinity of a low-mass YSO. The paper is organized as follows. In Section \ref{smarts}, we present [Ne {\sc iii}] $\lambda3869$ detection in low-dispersion blue ($3500\lesssim\lambda/{\rm \AA}\lesssim5500$) SMARTS/CTIO spectra of Sz 102. The rest of the paper is devoted to studies in the optical wavelengths from archival high-dispersion spectra obtained with the VLT. Basic properties and reduction of the two-dimensional optical spectra are presented in Section \ref{data_reduction}. In Section \ref{results}, [Ne {\sc iii}] $\lambda3869$ and selected forbidden line profiles are analyzed, and physical conditions from optical forbidden line ratios are derived. Based on the information obtained from multiple forbidden emission lines as well as the optical [Ne {\sc iii}] $\lambda3869$ line itself, we discuss origins of the emissions and physical conditions of the microjets. We also explore scenarios that could lead to the irradiation by X-rays in the Sz 102 system in Section \ref{discussion} and summarize the perspectives in Section \ref{summary}.
     
\section{REDUCTION AND ANALYSIS OF LOW-DISPERSION BLUE SPECTRA}\label{smarts}

\begin{figure}
\epsscale{0.8}
\plotone{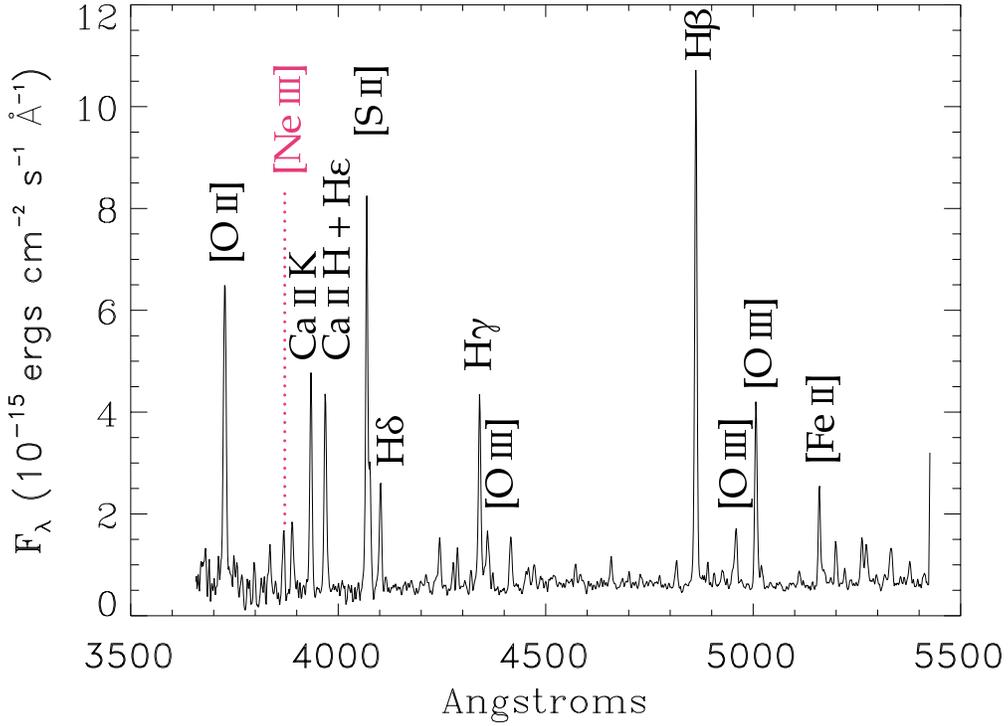}
\epsscale{1.0}
\caption{The low-dispersion blue spectrum of Sz~102 obtained with the SMARTS/CTIO RC spectrograph on 2011 July 21. The spectrum has been smoothed with a Fourier filter. Position of the [Ne {\sc iii}] $\lambda3869$ detection is denoted by a vertical dotted line. A more complete list of line detections is given in Table \ref{tbl-blow}.}
\label{fig-blow}
\end{figure}

      On 2011 July 21 we obtained a low-dispersion ($R\sim1000$) blue spectrum of Sz~102 (Figure~\ref{fig-blow}). The data were obtained using the SMARTS/CTIO 1.5m with the long slit RC spectrograph in first order with grating 26 \citep[see][for a description of the instrumental setup and data reduction]{Wal12}. The reduced image is the sum of three 1200 s integrations. The slit width is 1\farcs0; the slit is oriented fixed at 90\degr. At the time of observation the air mass was 1.03, the parallactic angle was 60\degr, and the differential offset between the blue spectrum and red guider image was of order 10\% of the seeing disk, and hence can be neglected. Flux calibration is accomplished by scaling from observations of the spectrophotometric standard LTT~4364.

      \cite{Krautter84} do not quote intensities for lines shortward of Ca~{\sc ii}~K, and do not show a plot of their blue spectrum. The most prominent lines in the blue spectral region in order of decreasing intensity are H$\beta$ $\lambda$4861, [S {\sc ii}] $\lambda\lambda4068+4076$, [O {\sc ii}] $\lambda\lambda3726+3729$, and [O {\sc iii}] $\lambda5007$. The upper Balmer lines are visible through H10 $\lambda3798$. H$\epsilon$ is abnormally strong due to blending with Ca~{\sc ii}~H. Otherwise, the strongest line in this region is [Ne {\sc iii}] $\lambda3869$. Note that the other half of the [Ne~{\sc iii}] doublet, at 3968\AA, is blended with Ca~{\sc ii}~H and H$\epsilon$.

       Line intensities, relative to H$\beta$, are given in Table~\ref{tbl-blow}. Intensities of the lines are not in great agreement with those of \citet{Krautter84}. This is likely attributable to the different slit sizes, and perhaps some stellar variability. In particular, the large discrepancy in the strengths of the 3933, 3968, and 4072 lines may be due to the fact that the SMARTS slit is relatively narrow, and fixed in position angle. We likely lost some blue light. \citet{Krautter84} comment on the  H$\gamma$/H$\beta$ ratio. They do not resolve H$\gamma$ from [O {\sc iii}] $\lambda4363$; the latter is about 25\% of the unresolved flux.  

\section{REDUCTION AND ANALYSIS OF ARCHIVAL VLT/UVES DATA} \label{data_reduction}
  
     The spectra of Sz 102 taken with the VLT/{\sc Uves} were obtained through the European Southern Observatory (ESO) archive \citep[][PI: F. Comer\'on]{CF10}. Two exposures, each of 1512 s, were made on 2003 June 9. The dichroic splits the spectra into wavelength ranges of 3300 -- 4500 \AA\ and 4800 -- 6800 \AA\ with slit lengths of 8\arcsec\ and 12\arcsec, respectively. The width of the slit was 1\farcs2, and the average resolution is $R\approx 33,000$ ($\Delta v \approx 10$ km\,s$^{-1}$). During the observations, the slit was kept parallel to the parallactic angle in order to minimize flux loss due to atmospheric scattering. This results in slit misalignment with the jet axis ranging from 4\degr\ to 18\degr\ during the observation. To mitigate the effect of misalignment, only the first exposure (with an average misalignment of $\sim 7\fdg5$) was used for analysis.

\begin{figure}
\plotone{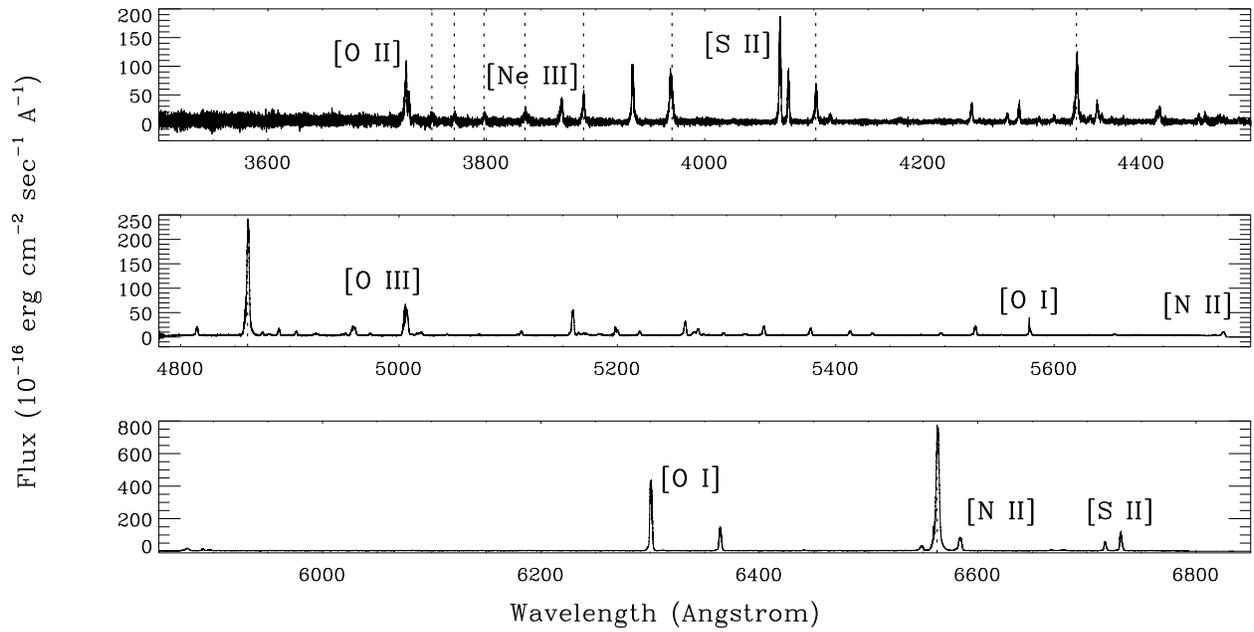}
\caption{The archival high-dispersion {\sc Uves} spectrum of Sz 102 obtained on 2003 June 9. Both the blueshifted and redshifted jets contribute to the emission. Forbidden emission lines used in this study are labeled by the species. Vertical dashed lines show the positions of the Balmer series up to H12.}
\label{1dspec}
\end{figure}

     Two-dimensional spectral images were reduced using the ESO {\sc Uves} pipeline version 4.9.8. Bias subtraction, order determination, flat-fielding, wavelength and flux calibration were executed during the pipeline cascade. Night-sky emission lines were identified based on the list of \citet{SkyLine}, and then spatially averaged and subtracted from the spectra. The heliocentric correction, obtained from the pipeline-reduced one-dimensional spectrum, was used to correct the wavelengths in the two-dimensional spectra. Furthermore, all the wavelengths were shifted by the $+4$ km\,s$^{-1}$ systemic velocity \citep{T01} of the star. Figure \ref{1dspec} shows the one-dimensional spectrum extracted from the flux-calibrated two-dimensional spectra.

     Only the reflection nebula associated with Sz 102, not the star, is visible in the optical. Its wavelength-dependent peak positions and widths were determined by fitting the spatial profiles with one-dimensional Gaussians. For every 100 pixels in the wavelength domain, spatial profiles in the line-free regions were binned and fitted. The FWHM of the spatial profiles decrease from 1\farcs5 at 3800 \AA\ to 1\farcs2 at 6300 \AA. The trend of each Gaussian parameters was modeled by a second-order polynomial. The modeled peak positions at different wavelengths are set as $y=0$.

      We analyzed the following forbidden emission lines: [O {\sc ii}] $\lambda\lambda3726,\,3729$, [Ne {\sc iii}] $\lambda3869$, [S {\sc ii}] $\lambda\lambda4068,\,4076$, [O {\sc iii}] $\lambda\lambda4959,\,5007$, [O {\sc i}] $\lambda5577$, [N {\sc ii}] $\lambda5755$, [O {\sc i}] $\lambda\lambda6300,\,6363$, [N {\sc ii}] $\lambda\lambda6548,\,6583$, and [S {\sc ii}] $\lambda\lambda6716,\,6731$. For each of the lines, the baselines of the two-dimensional spectra were subtracted by fitting row-by-row with a second-order polynomial within the velocity range of $-700$ to $+700$ km\,s$^{-1}$. The baseline-subtracted two-dimensional spectra were binned to the velocity resolution of 10 km\,s$^{-1}$ to increase the signal-to-noise ratio. 

     The binned two-dimensional spectra were analyzed by fitting one-dimensional Gaussians along both the spatial and dispersion axes. Offsets from the centroid of reflection nebula ($y=0$) were determined by fitting a single-component Gaussian along the cross-dispersion axis for each spectral bin. For each spatial row, the line profile was fitted with multiple Gaussians, depending on the skewness. Based on the peak positions of the Gaussian fitting, each row was decomposed into a blueshifted and a redshifted component by subtracting either the redshifted or the blueshifted Gaussians. For each of the lines, a set of ``blue'' spectra were then formed from the residual blueshifted component, and a set of ``red'' spectra were formed from the redshifted component. For presentation purposes, the spectra were smoothed with a two-dimensional Gaussian with $\sigma_G = 1$ pixel.

\section{RESULTS} \label{results}

\subsection{Forbidden Line Properties of Sz 102}

\begin{figure}
\epsscale{0.8}
\plotone{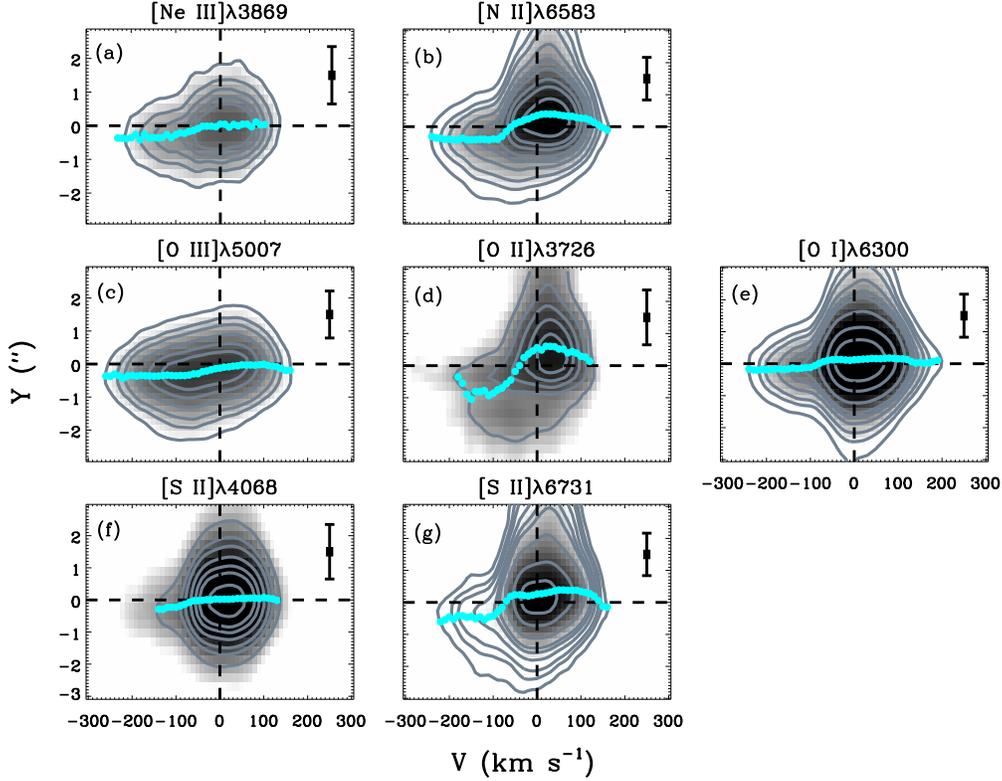}
\epsscale{1.0}
\caption{(a) -- (g) Position--Velocity (PV) diagrams of selected optical forbidden emission lines from Sz 102. All the spectra have been binned every 10 km\,s$^{-1}$ and smoothed with Gaussian with $\sigma_G = 1.0$ pixel. The velocities are relative to the systemic value. The positions are along the position angle of $\sim 109\degr$ and are relative to the fitted positions of point spread functions at each wavelength. The FWHM resolutions of velocity ($\sim 10$ km\,s$^{-1}$) and position (varying at different wavelengths) are shown at the upper-right corner of each panel. The filled circles connected across different velocity bins are fitted spatial centroids along velocity channels. Logarithmic gray scales are shown in all panels from $5.0\times10^{-17}$ to $1.0\times10^{-15}$ erg\,s$^{-1}$\,cm$^{-2}$\,\AA$^{-1}$. The noise levels and contour values of the lines are summarized in Table \ref{LineProperties}.}
\label{FELPVs_SpatCents}
\end{figure}

      Position--velocity (PV) diagrams of various forbidden lines are shown in Figure \ref{FELPVs_SpatCents}. Properties of the emission lines, including ionization potentials (I.P.) of the species, critical densities ($n_{\rm cr}$), upper-level energies relative to the ground state ($E_u$), and noise and contour levels of PV diagrams, are summarized in Table \ref{LineProperties}. The panels are arranged such that the ionization potentials generally decrease from left to right and from top to bottom, and critical densities decrease from left to right for the same ionization state ([S {\sc ii}]) and from top to bottom for the central column ([N {\sc ii}], [O {\sc ii}], and [S {\sc ii}]). Cyan filled circles overlaid on the PV diagrams are the fitted spatial centroids of different velocity bins. The centroid positions depend primarily on $n_{\rm cr}$; line species with $n_{\rm cr} > 10^5$ cm$^{-3}$ tend to have centroids within 0\farcs3 from the position of reflection nebula but those with $n_{\rm cr} < 10^5$ cm$^{-3}$ (the central column) have centroids at up to or beyond 0\farcs5. A trend that the redshifted spatial centroids are closer to $y=0$ than the blueshifted counterpart can be seen in high-ionization lines. The trend can also be seen but is less obvious in low-density lines. Such relative spatial offsets seem to be counter-intuitive for an edge-on system with a flat disk that usually has its reflection nebula at the blueshifted side. The true stellar position may be anywhere between the positions of the blueshifted and redshifted peaks and cannot be determined from the current data set. The spatial extension of the emission, especially for the redshifted side, is larger for species with either low I.P. or low $n_{\rm cr}$. The two high-I.P. species, [Ne {\sc iii}] and [O {\sc iii}], are confined within $\pm2\arcsec$. [S {\sc ii}] $\lambda4068$ is spatially more confined than [S {\sc ii}] $\lambda6731$, whose critical density is about three orders of magnitude lower.

\begin{figure}
\epsscale{0.8}
\plotone{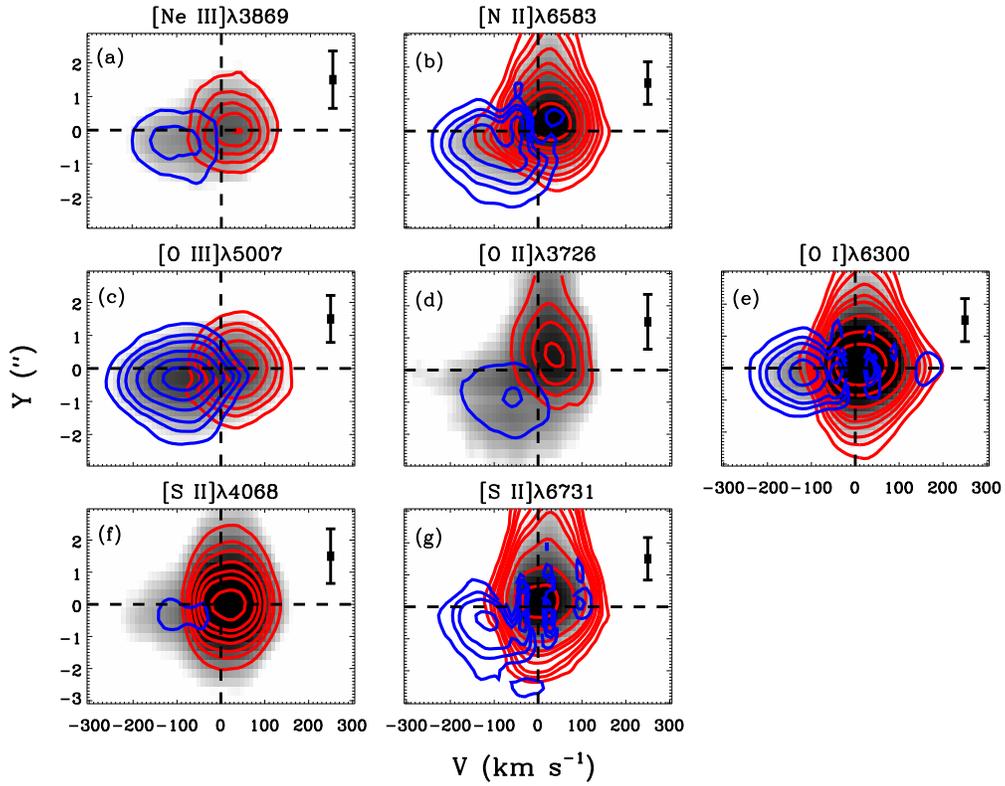}
\epsscale{1.0}
\caption{PV diagrams of forbidden lines as listed in Figure \ref{FELPVs_SpatCents}. Shown in grey scales are spectra directly extracted from the data as in Figure \ref{FELPVs_SpatCents}, and shown in blue and red contours are decoupled spectra after velocity decomposition using multiple Gaussian fitting. The blue contours seen in the redshifted velocities are subtraction residuals from velocity decompositions, which is most evident in panels (b), (e), and (g).}
\label{FELPVs_VelCents}
\end{figure}

      In Figure \ref{FELPVs_VelCents} the PV diagrams are shown in background gray scales as in Figure \ref{FELPVs_SpatCents} and those from the decomposed blue and red spectra are displayed in blue and red contours, respectively. The decomposed spectra show that the overall shape of the spectra is due to spatial blending caused by seeing $\sim1\farcs5$. Therefore, the kinematics of the lines are better deciphered after decomposition. For most of the lines, the redshifted component is stronger and centered at a velocity $\sim+25$ km\,s$^{-1}$, with velocity widths ranging from 100 to 140 km\,s$^{-1}$. This redshifted emission, extending beyond 4\arcsec\ in low-I.P. species such as [N {\sc ii}], [O {\sc i}], [O {\sc ii}], and [S {\sc ii}], traces the receding jet \citep{GH88,CF10}. The kinematics traced by the blueshifted emission is remarkably different from its redshifted counterpart. The peaks of the lines range between $-120$ to $-90$ km\,s$^{-1}$, with an averaged line width of $\sim150$ km\,s$^{-1}$. The blueshifted peak velocities obtained from the line profiles appear to be a factor of $\sim1.5$ larger than the velocities obtained through low- to medium-dispersion spectra ($\sim-80$ km\,s$^{-1}$; \citealt{GH88}; $\sim-60$ km\,s$^{-1}$; \citealt{Cof07}).

      Asymmetries between the blueshifted and redshifted jets in velocity and intensity can also be seen among different emission lines. The most obvious difference appears in oxygen lines with different ionization states. In [O {\sc iii}] $\lambda\lambda4959,5007$ lines, the peak intensity and total flux from both sides are nearly identical. In [O {\sc ii}] $\lambda3726,3729$ lines, the blueshifted peak becomes $\sim0.4$ of the redshifted peak, and in [O {\sc i}] lines, the ratio drops to $\sim0.1$. For other lines, the ratios decrease from $\sim0.5$ in [Ne {\sc iii}], to $\sim0.2$ in [N {\sc ii}], and $\sim0.1$ in [S {\sc ii}] lines. That the trend of intensity contrast generally follows the ionization states of the lines suggests very different ionization conditions inside the two jets. The higher flux ratios in the high-excitation oxygen lines suggest that the ionization fraction in the blue jet is higher, despite the much weaker intensity due to the lower density overall.
      
        Emission spectrally extending across the systemic velocity was observed in both the blueshifted and redshifted jets in all the lines. The redshifted profile extends blueward up to $\sim-100$ km\,s$^{-1}$ and the blueshifted profile has a redward extension up to $\sim+100$ km\,s$^{-1}$. This kind of spectral feature was first mentioned as the ``excess'' emission by \citet{Pyo06} in comparing the observed [Fe {\sc ii}] spectra of the RW Aur A jet with the synthetic spectra predicted by models \citep[e.g.,][]{Shang98}. Observationally, it is easily missed after the continuum subtraction due to its relative weakness and its proximity to the scattered stellar continuum in some low-mass sources \citep[see discussions in, e.g.,][]{Pyo06,HH09,LS12}. It may be understood that for the case of Sz 102, its edge-on disk has reduced the amount of scattered stellar continuum and helps to raise the contrast between the jet emission and stellar contributions. This also results in fewer artifacts that may be introduced by the stellar continuum subtraction and enables a better assessment of the origins of emission lines since the entirety of the line profiles can be preserved. These features are prominent in all the forbidden lines reported here, especially in the redshifted jet.
        
       The high-dispersion {\sc Uves} spectra can be compared with high-spatial-resolution {\it Hubble Space Telescope} Imaging Spectrograph ({\it HST}/STIS) spectra observed 0\farcs3 from the source at 0\farcs1 and 65-km\,s$^{-1}$ resolutions \citep{Cof07,Cof12}. The line profiles on the receding side consist of a peak at $\sim +55$ km\,s$^{-1}$ with a shoulder extending blueward up to $-120$ km\,s$^{-1}$; those on the approaching side have a large width spanning from $\sim-200$ to $\sim+200$ km\,s$^{-1}$ \citep[see Figure 3 of][]{Cof12}. Comparing with the {\sc Uves} spectra that were resampled and convolved with the STIS instrumental profile, the somewhat larger line widths and higher peak velocities of the redshifted emission in the STIS spectra may be an effect of its lower velocity resolution and lower spectral sampling rate. Comparisons for the blueshifted line profiles are less robust since both spectra are faint. Imperfect decomposition residuals in the blueshifted components of the {\sc Uves} spectra may also cause confusion. The overall agreements from the two data sets suggest that the {\sc Uves} spectra mainly trace the jet emission from a nearly edge-on system and that the velocity decomposition can separate the two jets despite confusion close to the star. 

\subsection{Optical [Ne {\sc iii}] Forbidden Line Emission from Sz 102}
 
\begin{figure}
\epsscale{0.8}
\plotone{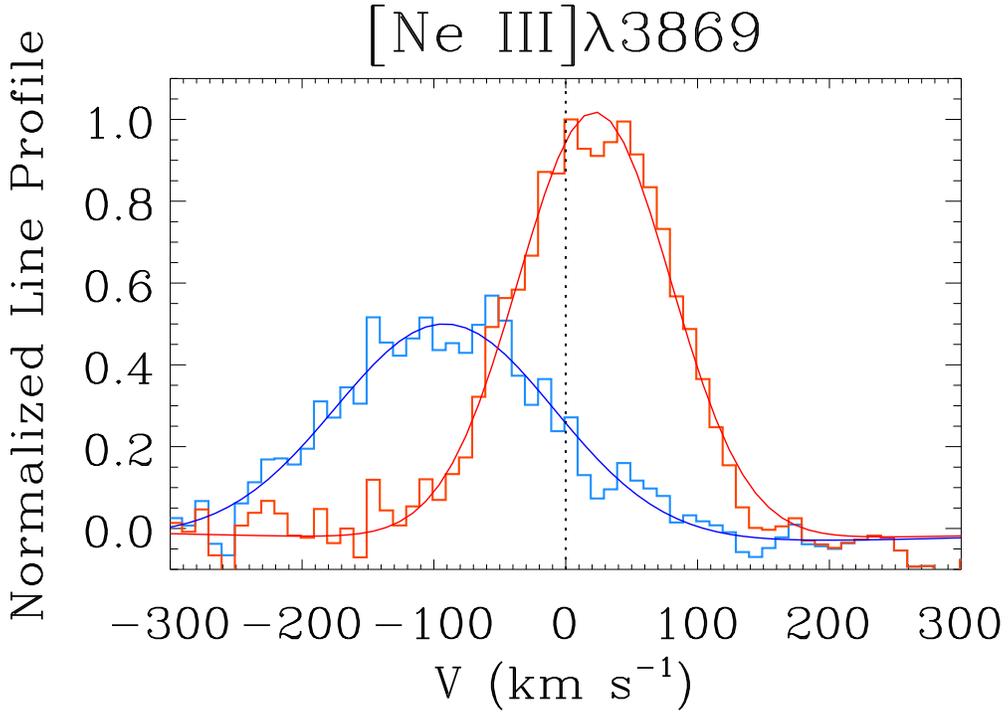}
\epsscale{1.0}
\caption{Decomposed [Ne {\sc iii}] $\lambda3869$ line profiles normalized to the redshifted peak intensity. The spectra have been binned every 10 km\,s$^{-1}$ in the spectral direction. The blueshifted and redshifted emission are shown in blue and red histograms, respectively. For each decomposed line profile, a single Gaussian is fitted and is shown in blue and red think lines, respectively. The fitted peak position and FWHM for the blueshifted emission are $-91\pm3$ km\,s$^{-1}$ and $196\pm9$ km\,s$^{-1}$, and for the redshifted emission are $+21\pm1$ km\,s$^{-1}$ and $135\pm2$ km\,s$^{-1}$.}  
\label{NeIIILP}
\end{figure}

      [Ne {\sc iii}] $\lambda3869$ stands out in the forbidden emission lines observed from Sz 102 as the species with the highest ionization potential and one whose critical density is among the highest. From the decomposed spectra, this doubly-ionized neon line is seen in both of the microjets, separated by $\sim0\farcs3$ by their respective centroids in the images. The line profiles of blueshifted and redshifted [Ne {\sc iii}] $\lambda3869$ emission are shown in Figure \ref{NeIIILP} by integrating the decomposed spectra along the spatial axis. The profiles are slightly skewed, with the blueshifted emission having a redshifted tail, and vice versa. By fitting individual profile with a Gaussian, the redshifted emission peaks at $+21\pm1$ km\,s$^{-1}$ with an FWHM of $135\pm2$ km\,s$^{-1}$, and the blueshifted emission peaks at $-91\pm3$ km\,s$^{-1}$ with an FWHM of $196\pm9$ km\,s$^{-1}$. The redshifted emission is stronger with a integrated flux (within $\pm2\,{\rm FWHM}$ of the seeing) of $5.0\times10^{-15}$ erg\,s$^{-1}$\,cm$^{-2}$ and the blueshifted emission has an integrated flux of $2.4\times10^{-15}$ erg\,s$^{-1}$\,cm$^{-2}$. 
   
\begin{figure}
\plottwo{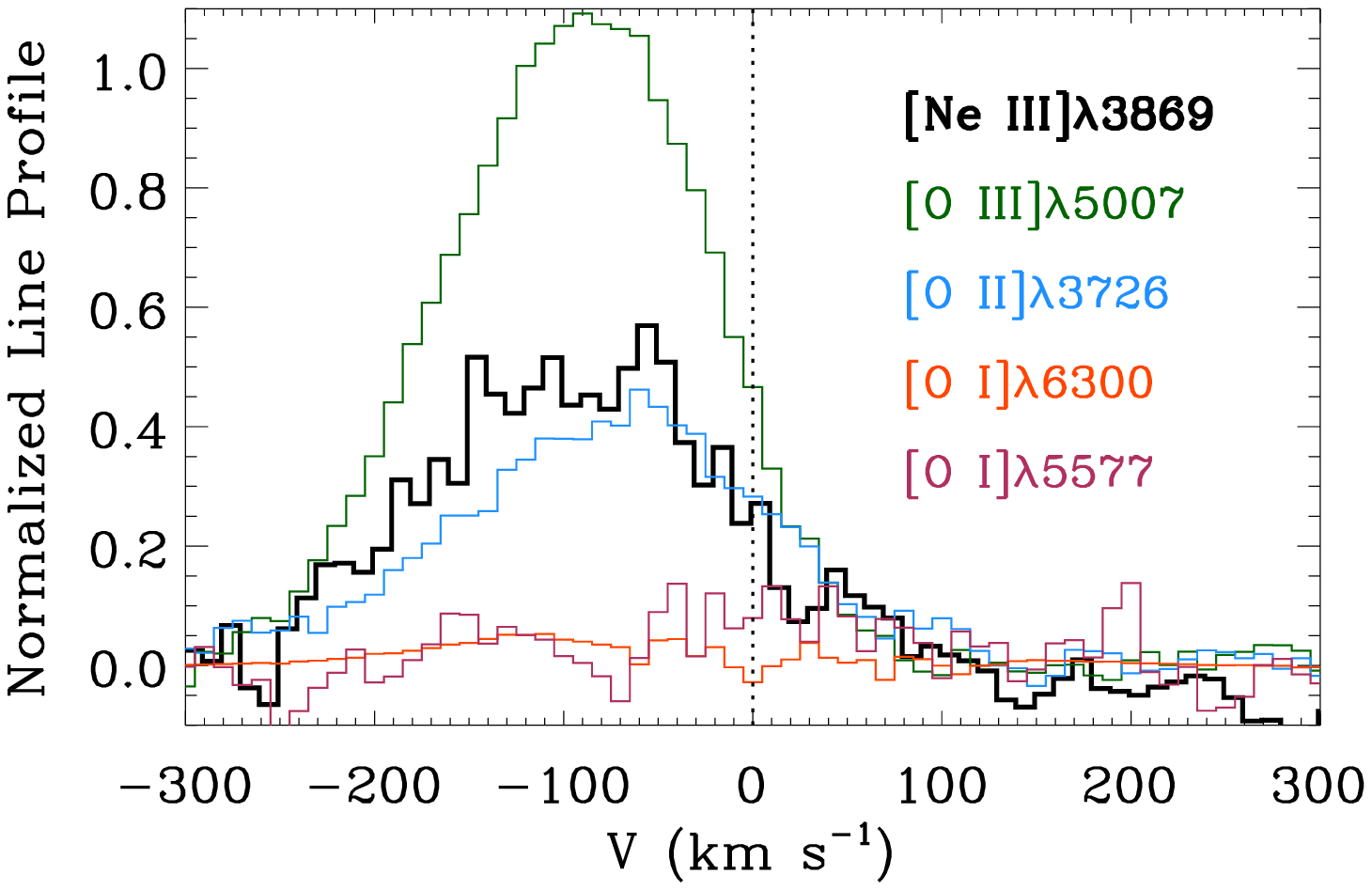}{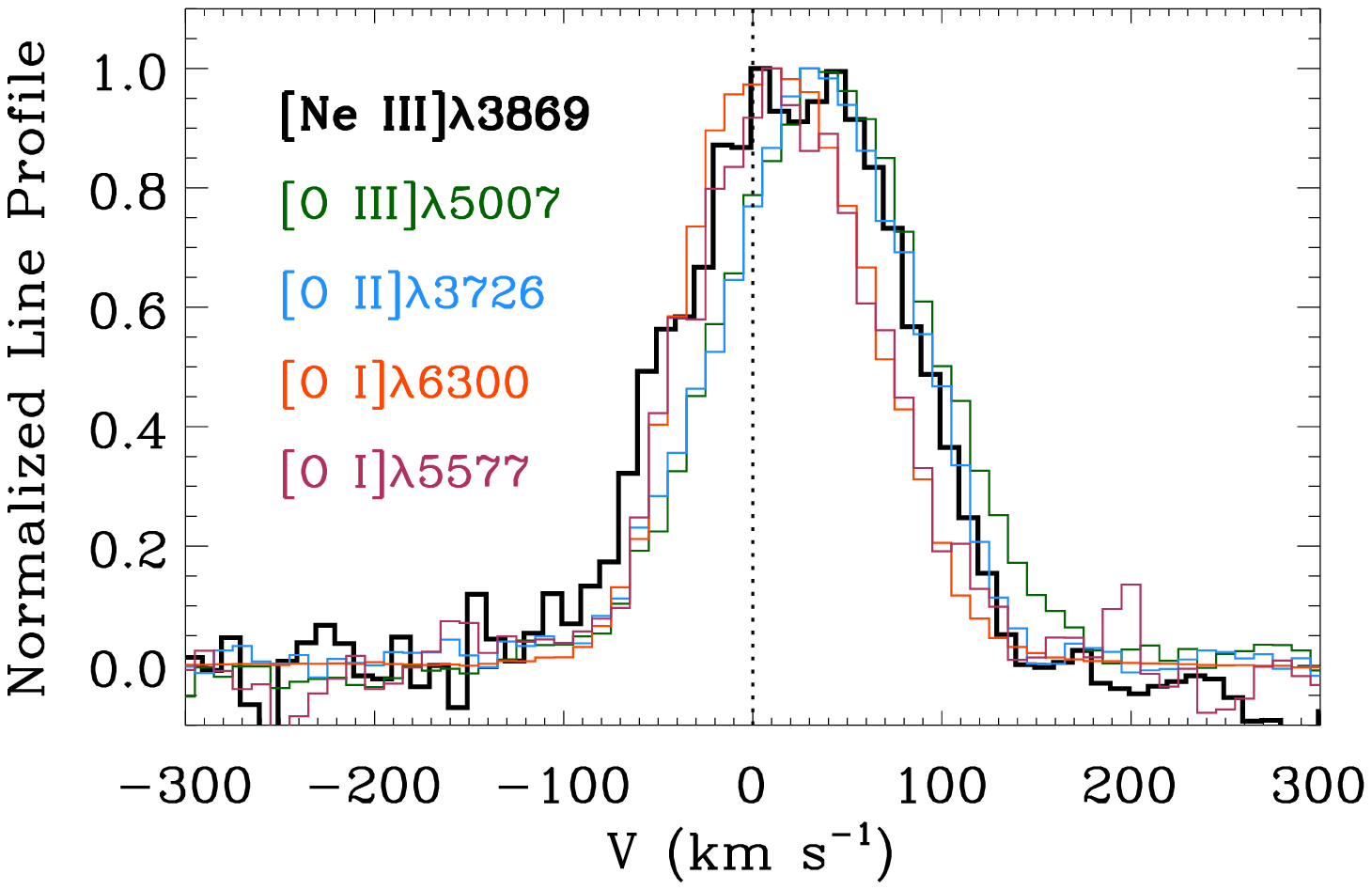} \\
\plottwo{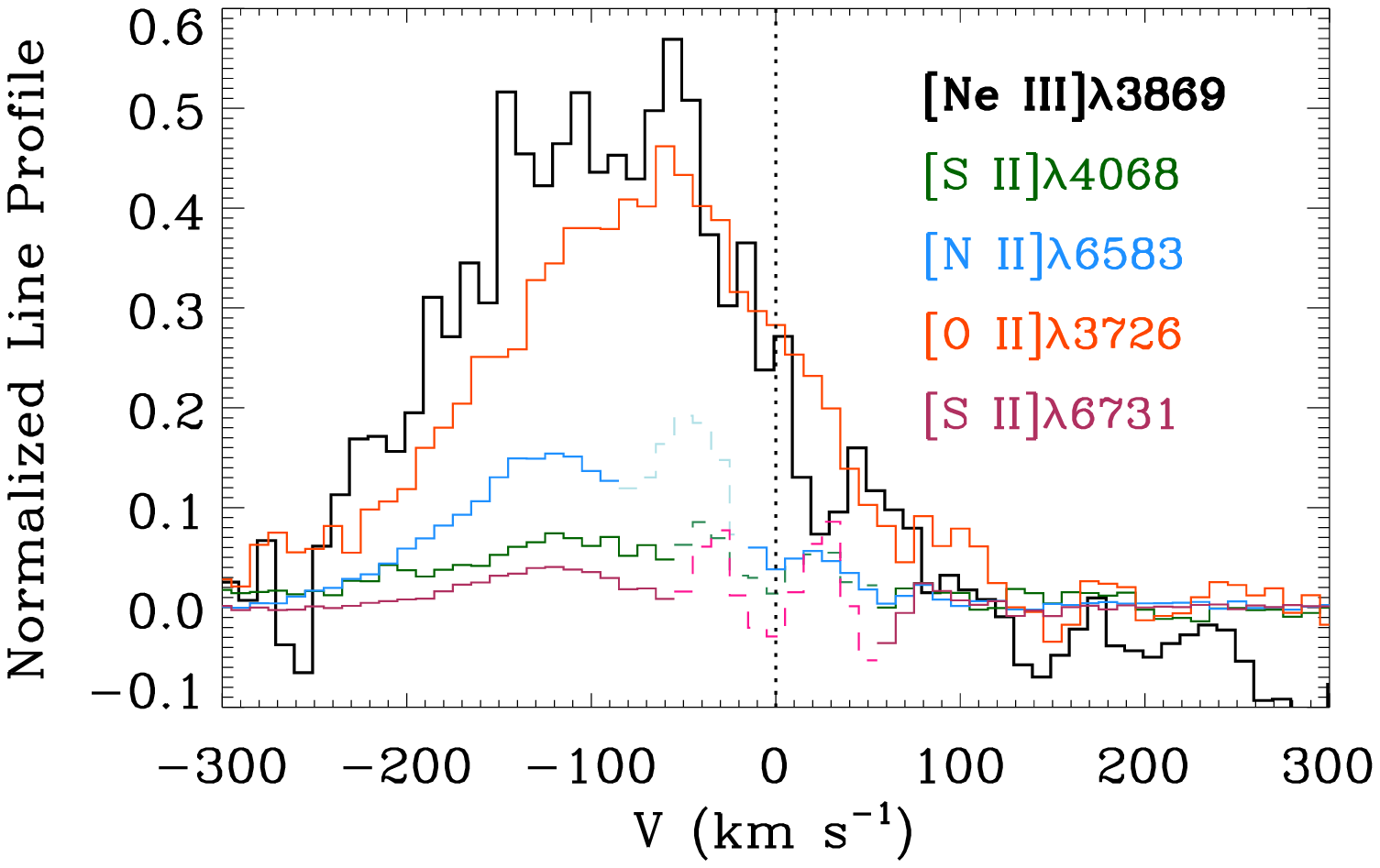}{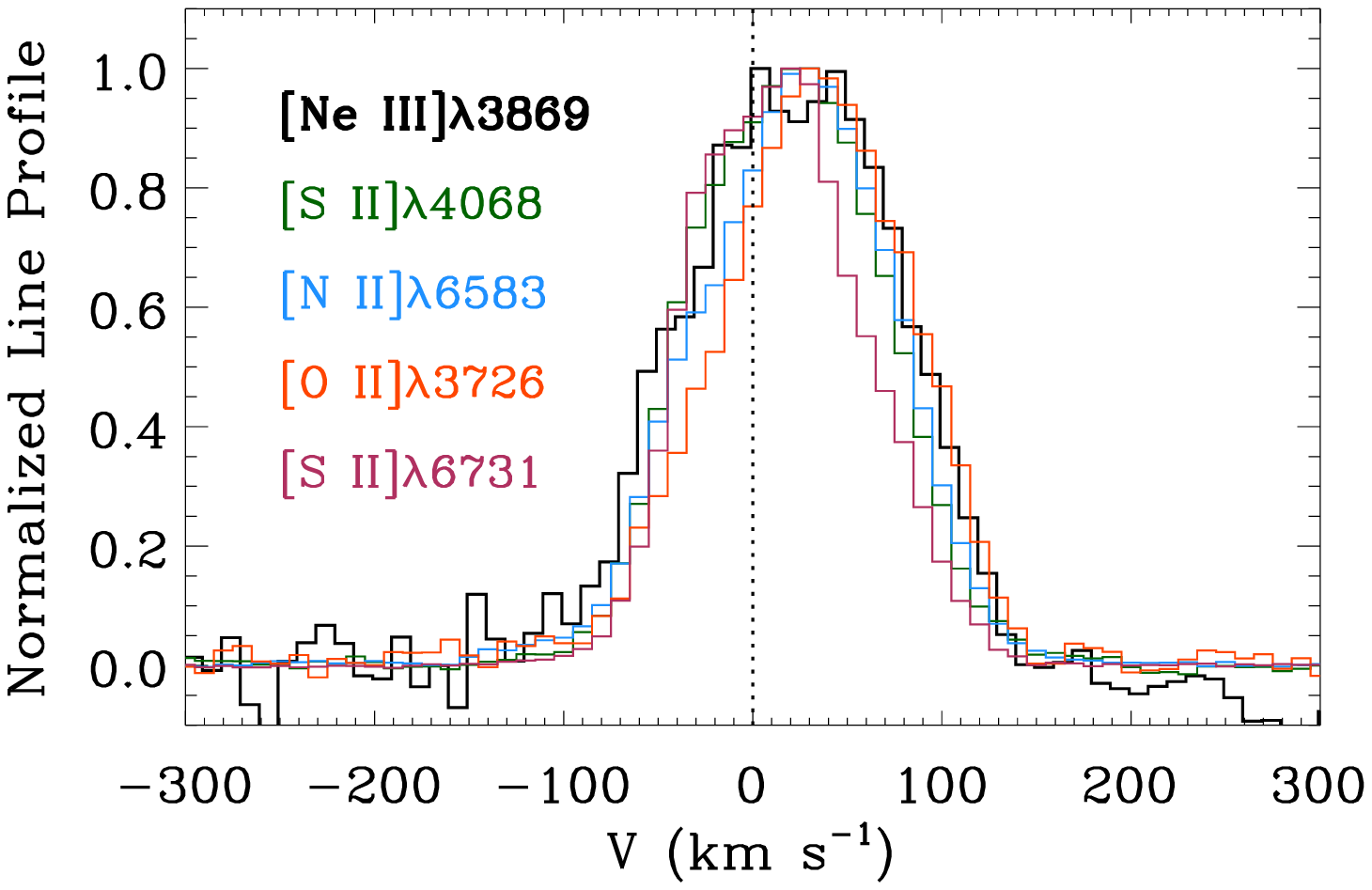}
\caption{Decomposed line profile comparisons of [Ne {\sc iii}] $\lambda3869$ with other forbidden lines. Comparisons on decoupled blueshifted (left column) and redshifted (right column) emission are shown with normalization to their individual redshifted peak intensities. Comparisons with with oxygen lines of different ionization states are shown in the upper row and comparisons with singly-ionized lines with different ionization potentials and critical densities are shown in the lower row. Vertical scales from $-0.1$ to $1.1$ are shown, except for the lower right panel where $-0.1$ to $0.6$ are shown. The regions of imperfect decomposition residuals in the redshifted [N {\sc ii}] and [S {\sc ii}] profiles have been marked out by dashed histograms.}
\label{LPCompare}
\end{figure}

     [Ne {\sc iii}] has the broadest profiles among the detected forbidden lines. The kinematic properties of [Ne {\sc iii}] $\lambda3869$ are compared to those of other forbidden lines through normalized line profiles as in Figure \ref{LPCompare}. The left column shows the decomposed blueshifted profiles and the right column shows the decomposed redshifted profiles, all normalized to their individual redshifted peaks. The upper row compares [Ne {\sc iii}] $\lambda3869$ with oxygen lines with descending ionization states, and the lower row compares [Ne {\sc iii}] $\lambda3869$ with low-ionization lines in the order of decreasing critical densities.  The blueshifted [Ne {\sc iii}] has its velocity peak at $\sim-100$ km\,s$^{-1}$, as most of the low-ionization lines [O {\sc i}], [N {\sc ii}], and [S {\sc ii}].
      
     Our spectrally-resolved [Ne {\sc iii}] $\lambda3869$ detection from Sz 102 demonstrates its important role in tracing the innermost region of the wind. A signature of an intensity dip was found at the peak position of the redshifted [Ne {\sc iii}] $\lambda3869$ profile (Figure \ref{NeIIILP}). The width of the dip is $\sim50$ km\,s$^{-1}$. The relative strength of the dip ($\sim 10\%$) compared with the noise levels ($\sim 4\%$) suggests that the dip may be an authentic spectral feature. Whether a dip exists in the blueshifted peak is difficult to judge due to its lower signal-to-noise. This is also not obvious in other forbidden lines. Unlike the dips seen close to the systemic velocity in the {\it HST}/STIS UV--optical spectra \citep{Cof12}, the origin of the possible dip at the redshifted terminal jet velocity may be related to the dense inner wind. One possibility is the self-absorption from the innermost densest wind that has been accelerated close to the terminal velocity. Another possibility may be a geometric effect of a dense wind encompassing an empty axial region. Possible explanations will be discussed in a future publication.

\subsection{Line Ratios and Derived Physical Conditions}
       
\begin{figure}
\begin{center}
\epsscale{0.5}
\plotone{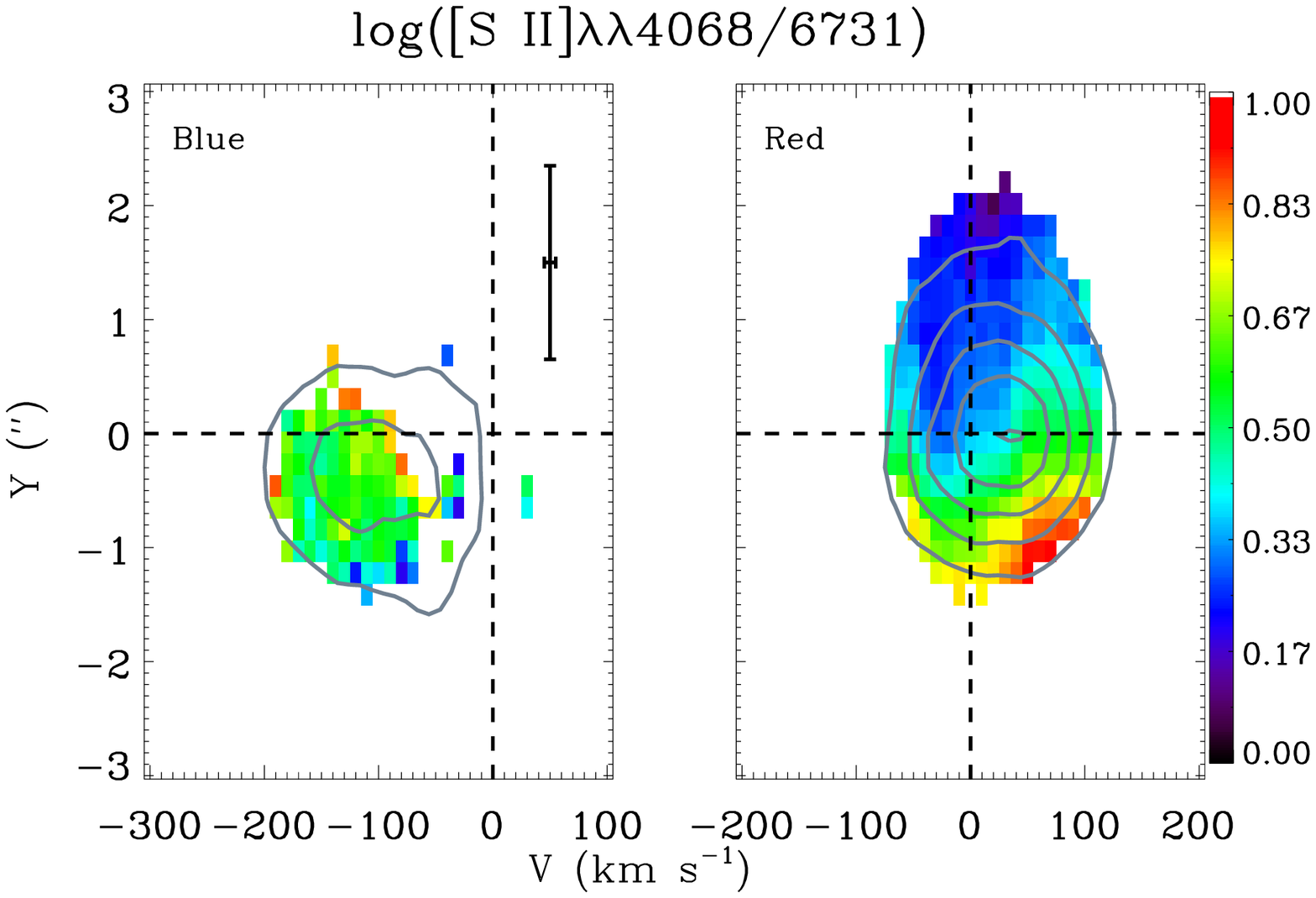} \\
\plotone{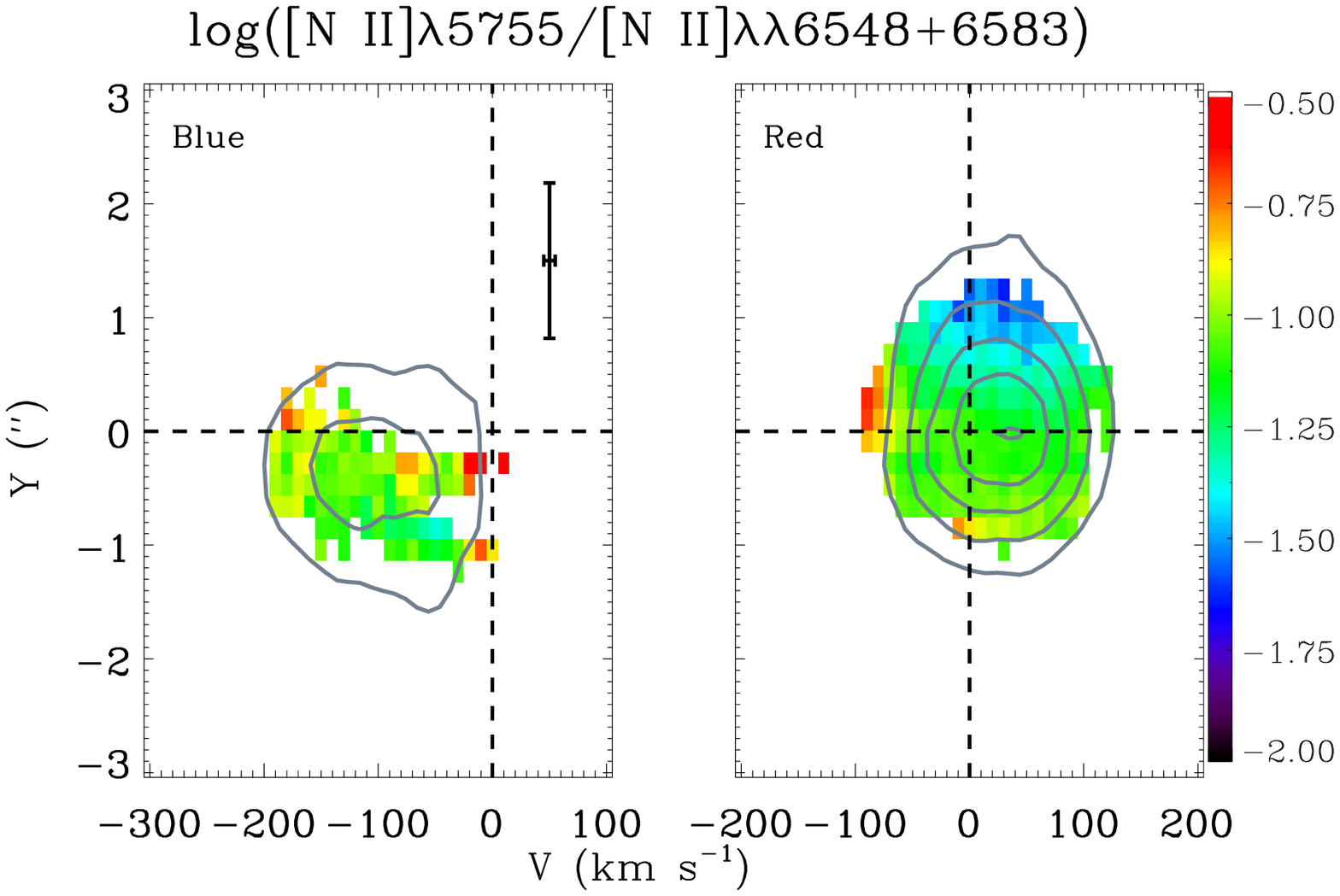} \\
\plotone{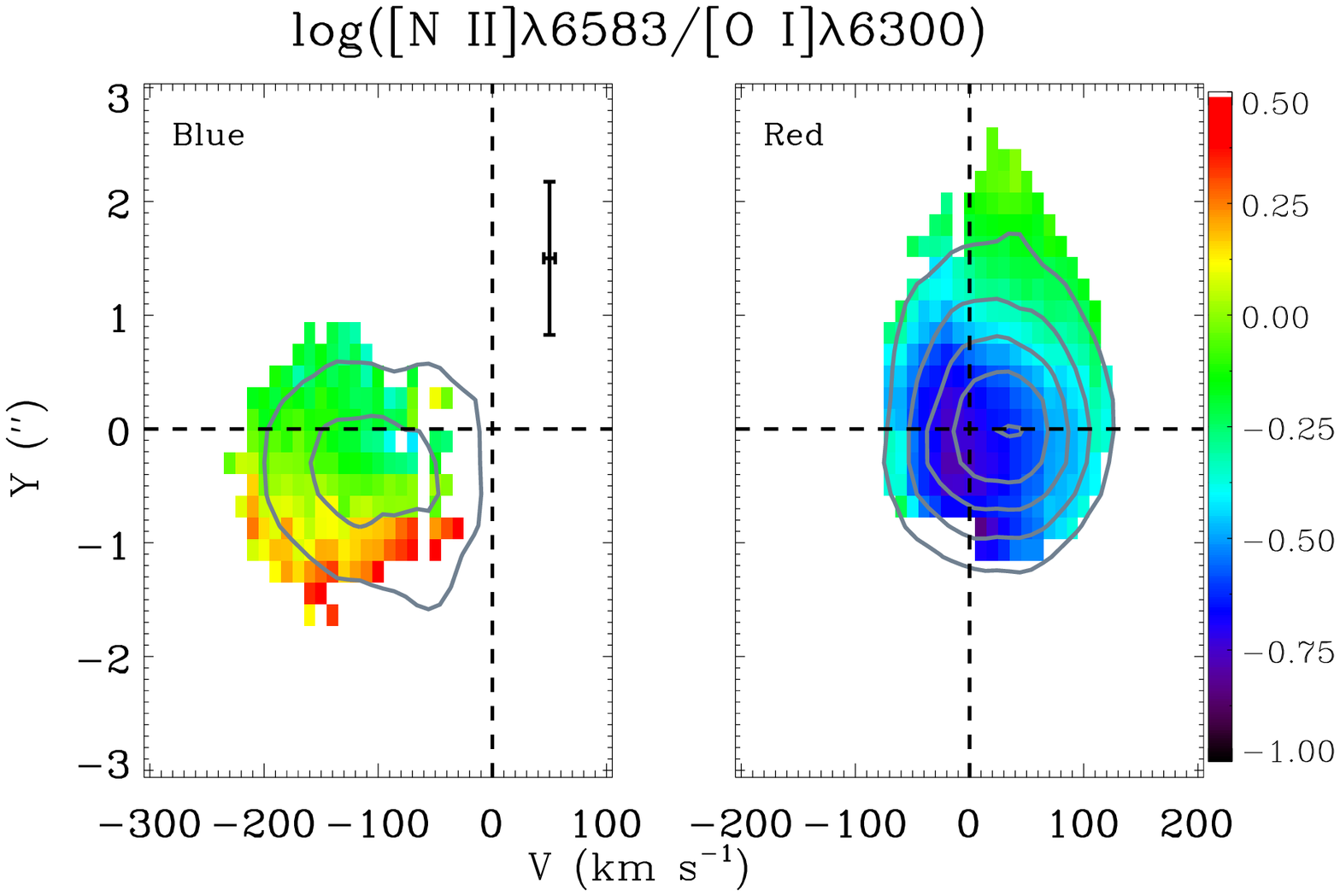}
\epsscale{1.0}
\caption{The line ratios of [S {\sc ii}] $\lambda\lambda4068/6731$ (top), [N {\sc ii}] $\lambda\lambda5755/6548+6583$ (center), and [N {\sc ii}] $\lambda6583/$[O {\sc i}] $\lambda6300$ (bottom) in the form of PV diagrams. The left panels are for the blueshifted jet and the right panels are for the redshifted jet. All values are in logarithmic scale. On each panel, velocity-decomposed [Ne {\sc iii}] $\lambda4869$ for the blueshifted and redshifted jets are overlaid with gray contours. The horizontal dashed lines represent the fitted position of the continuum and the vertical dashed lines represent the systemic velocity, as in Figure \ref{FELPVs_SpatCents}. For each line ratio, the velocity and spatial resolutions are plotted as error bars in the left (blueshifted) panels. The vertical error bars show the average FWHM of the PSF at the wavelength range of the respective line ratio, and the horizontal error bars show the velocity resolution of $\sim10$ km\,s$^{-1}$.}
\label{LineRatios}
\end{center}
\end{figure}

        Emission line ratios provide diagnostics for physical conditions of the jet. Figure \ref{LineRatios} shows the logarithmic line ratios of [S {\sc ii}] $\lambda\lambda4068/6731$, [N {\sc ii}] $\lambda5755/\lambda\lambda6548+6583$, and [N {\sc ii}] $\lambda6583/$[O {\sc i}] $\lambda6300$ (hereafter [S {\sc ii}] ratio, [N {\sc ii}] ratio, and [N {\sc ii}]/[O {\sc i}]), in terms of PV diagrams. Decomposed blueshifted and redshifted jet emission are plotted separately. The three line ratios are chosen because they are positively correlated with electron density $n_e$, electron temperature $T$, and ionization fraction $x_e$, respectively. To have better constraints in $n_e$ with a higher signal-to-noise, the stronger transitions of the [S {\sc ii}] $^2{\rm P}^\circ$ ($\lambda4068$) and $^2{\rm D}^\circ$ ($\lambda6731$) levels are used for the ratio. Although the temperature-sensitive ratio [Ne {\sc iii}] $\lambda\lambda3869+3968/\lambda3343$ cannot be obtained due to high background noise at $\lambda3343$ and line blending at $\lambda3968$, the high critical density of [N {\sc ii}] $\lambda5755$ ($\sim10^7$ cm$^{-3}$) can provide temperature estimates to high-density regions. Contours of [Ne {\sc iii}] $\lambda3869$ from the blueshifted and redshifted jets are plotted for reference. The presented line ratios are clipped by their relative errors such that they are within the spatial extent of $10\sigma$ level on the [Ne {\sc iii}] map. The general differences between the blueshifted and redshifted jets can be seen from the diagrams. The [S {\sc ii}] ratio is $\sim4$ within the blueshifted jet; in the redshifted jet the ratio starts at $\sim4$ close to the peak position of [Ne {\sc iii}] and decreases down to $\sim2$ along the flow. The [N {\sc ii}] ratio is higher in the blueshifted jet, $\sim0.1$ close to the peak, and slightly lower in the redshifted jet with a median value of $\sim0.07$. The [N {\sc ii}]/[O {\sc i}] ratio is $\sim1$ in the blueshifted jet. This ratio is $\sim0.3$ at the peak of redshifted emission and increases up to $\sim0.6$ due to further recombination in the flow.

\begin{figure}
\begin{center}
\plotone{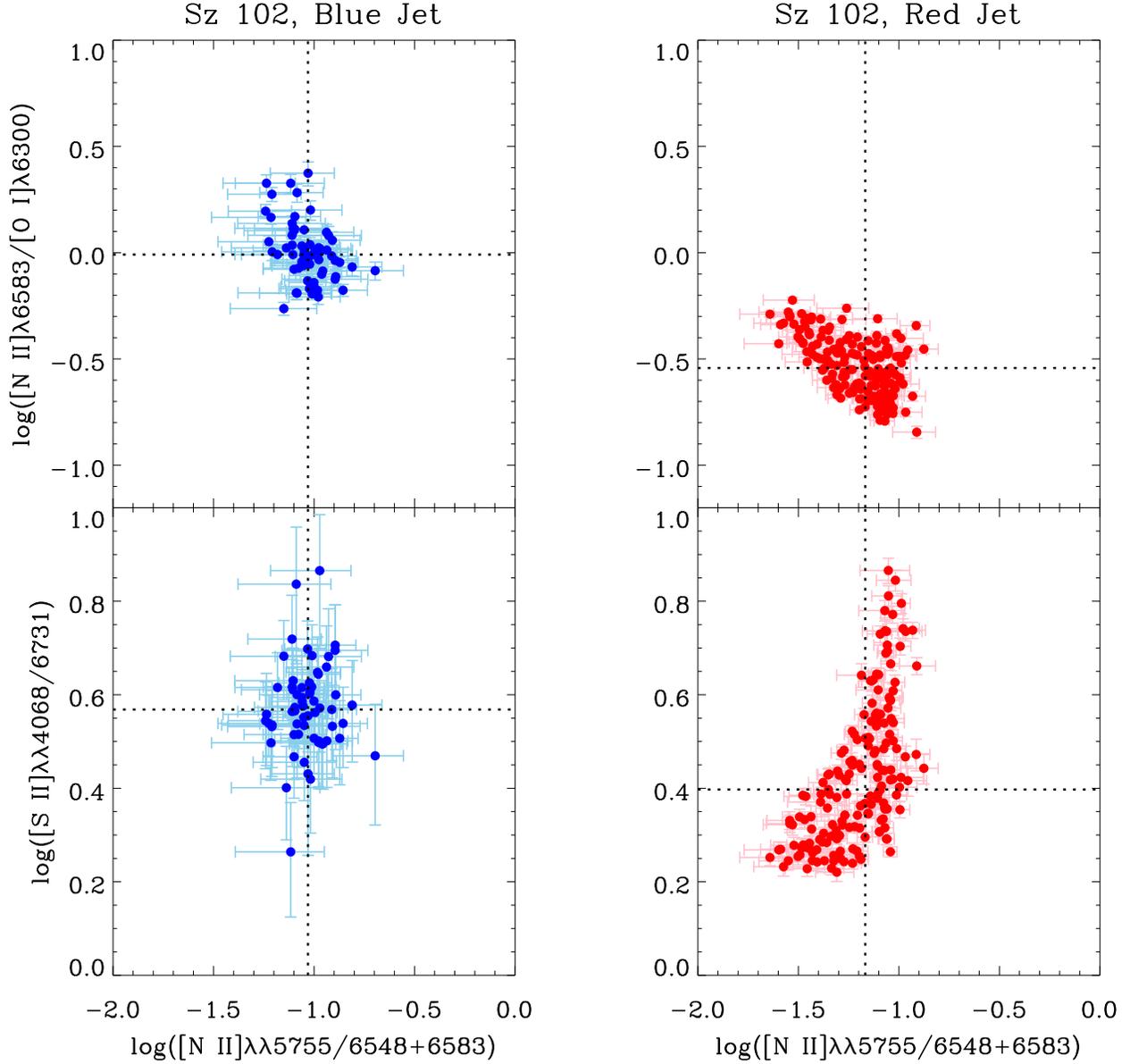}
\caption{Line-ratio diagnostic plots for the blueshifted (left panels) and redshifted (right panels) jets. The upper two panels show [N {\sc ii}] $\lambda6583/$[O {\sc i}] $\lambda6300$ against [N {\sc ii}] $\lambda\lambda5755/6548+6583$, and for the lower two panels the y-axes are replaced by [S {\sc ii}] $\lambda\lambda4068/6731$. Points are representative as those in Figure \ref{LineRatios}. Median values of each line ratio in each jet are shown in dotted lines. The median values for the blueshifted and redshifted jets are [N {\sc ii}] ratio: (0.09,0.07), [N {\sc ii}]/[O {\sc i}]: (0.98,0.29), and [S {\sc ii}] ratio: (3.70,2.52).}
\label{LRPlanes}
\end{center}
\end{figure}

        The differences of line ratios, as well as the inferred physical conditions, in the two jets can be seen more readily in Figure \ref{LRPlanes}. The temperature indicator, [N {\sc ii}] ratio, is shown on the {\sl x}-axis. [N {\sc ii}]/[O {\sc i}], representing $x_e$, and [S {\sc ii}] ratio, indicating $n_e$, are shown on the {\sl y}-axis in the upper and lower panels, respectively. Median values of the ratios from the individual jets are indicated by dashed lines. The most evident differences are shown in the panels between the [N {\sc ii}] ratio and the [N {\sc ii}]/[O {\sc i}] ratio. Loci of cross-ratio points from the blueshifted jet are clustered around [N {\sc ii}] ratio $\sim 0.1$ and [N {\sc ii}]/[O {\sc i}] $\sim 1$. Those from the redshifted jet are distributed from 0.03 to 0.1 and from 0.2 to 0.5. The [N {\sc ii}]/[O {\sc i}] ratios of the two jets do not overlap on the diagnostic plots although they follow a similar slope due to their different median values. A negative correlation in the upper panels may be a density effect; the drop-off of density (points with lower [S {\sc ii}] ratios) along the jet may produce lower [O {\sc i}] intensities and higher [N {\sc ii}]/[O {\sc i}] ratios. Data points on the panels between the [N {\sc ii}] ratio and the [S {\sc ii}] ratio are less well separated. There is a larger gradient in physical conditions in the redshifted jet, which gives a larger span of the [S {\sc ii}] and the [N {\sc ii}] ratios on the plot. Overall the median [S {\sc ii}] ratio is higher in the blueshifted jet than that in the redshifted jet by $\sim0.2$ dex.

\begin{figure}
\begin{center}
\epsscale{0.6}
\plotone{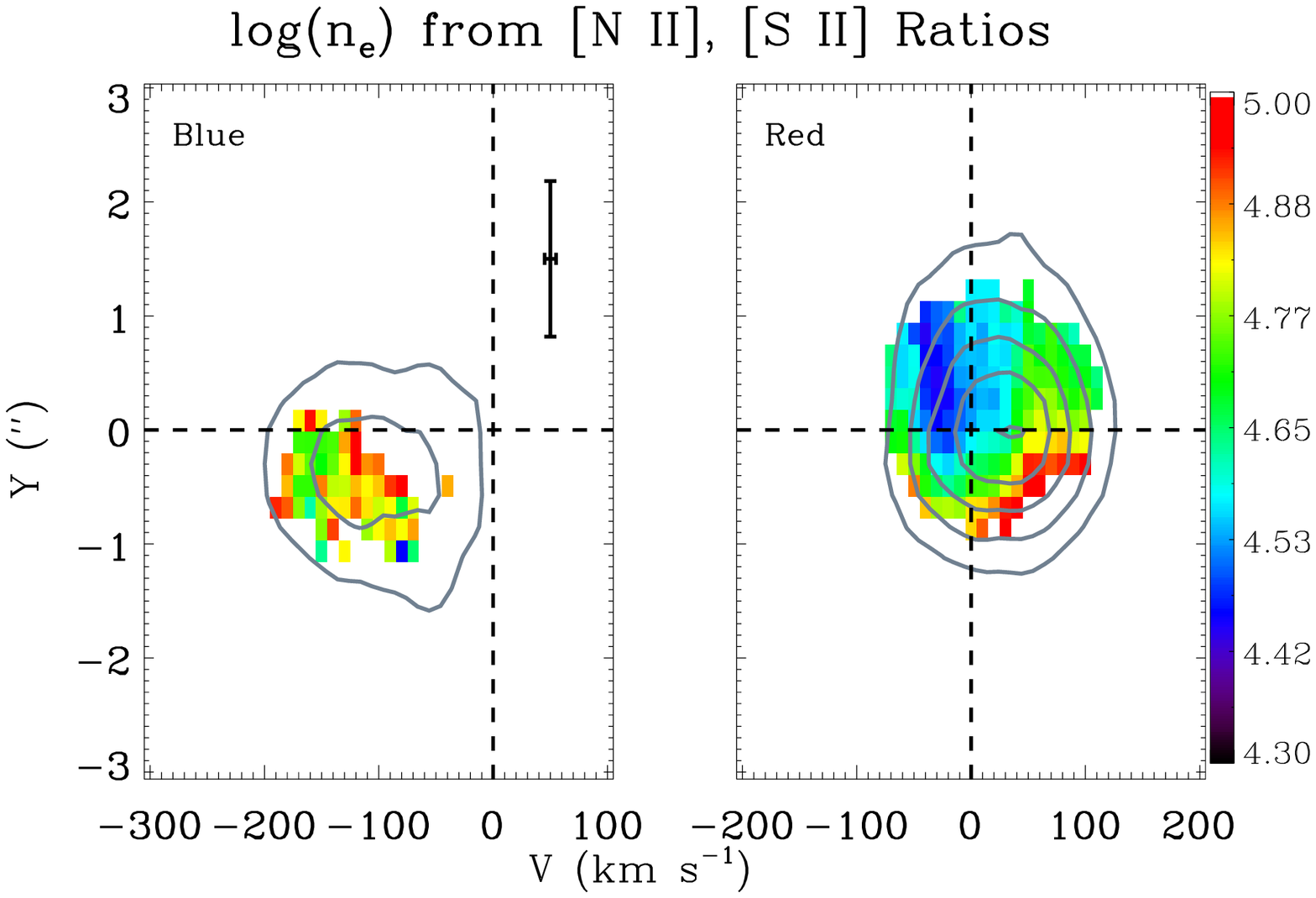} \\
\plotone{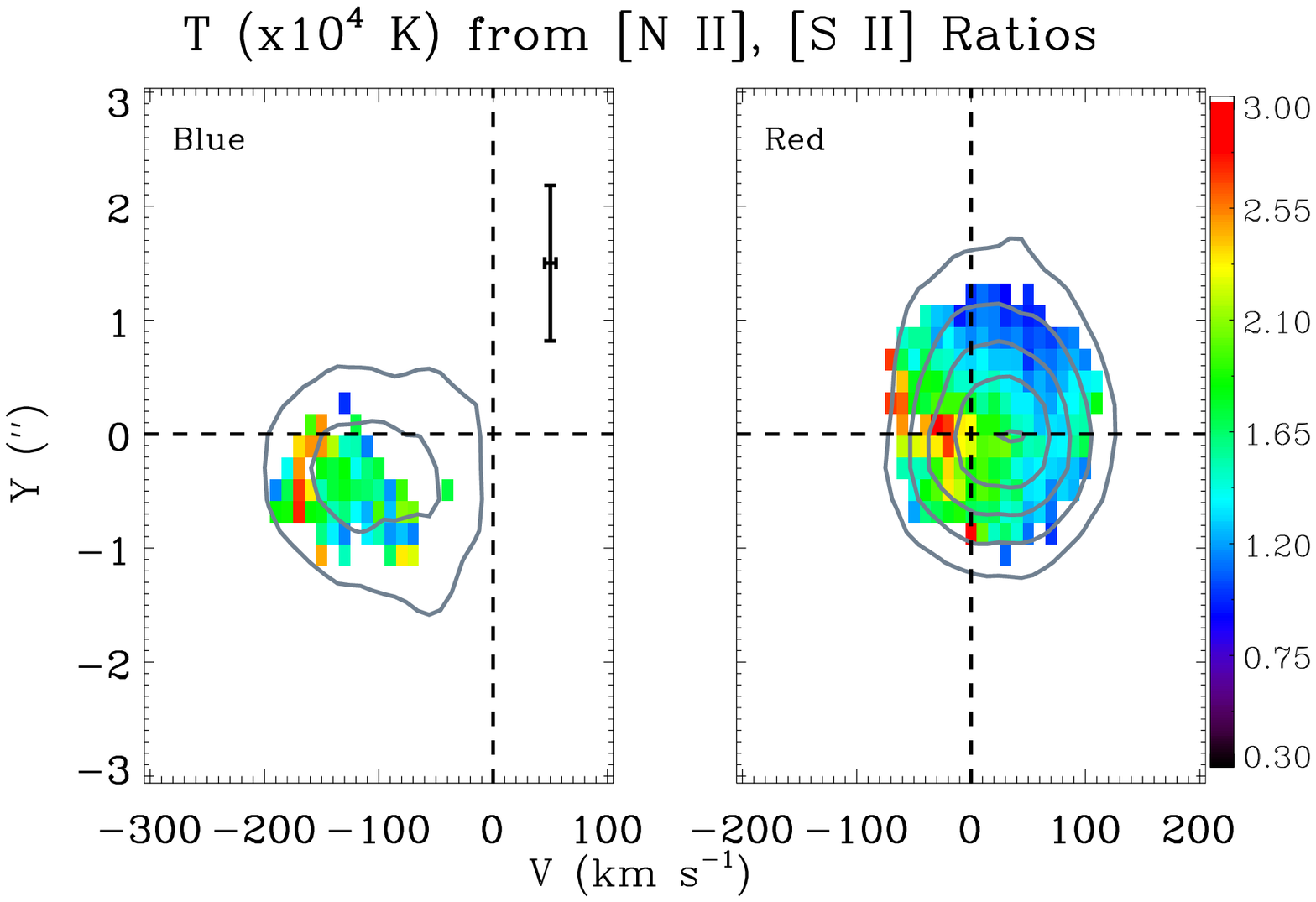}
\epsscale{1.0}
\caption{Derived electron densities ($n_e$, in logarithmic scales with units of cm$^{-3}$) and temperatures ($T$, in linear scales with units of $10^4$ K) derived from [S {\sc ii}] $\lambda\lambda4068/6731$ and [N {\sc ii}] $\lambda\lambda5755/(6548+6583)$ ratios. The values on each pixel is obtained by matching the two ratios on the ($n_e,T$) plane. The blueshifted jet is shown on the left panel and the redshifted jet is shown on the right panel. Velocity-decomposed [Ne {\sc iii}] $\lambda3869$ are plotted as gray contours. The dashed lines and error bars are of the same meanings as those in Figure \ref{LineRatios}.}
\label{NTPlots}
\end{center}
\end{figure}

          Figure \ref{NTPlots} presents the electron density and temperature distribution inferred from the [N {\sc ii}] and [S {\sc ii}] line ratios using formulae and atomic properties adopted by \citet{AGN2}. For each pixel, the observed [N {\sc ii}] and [S {\sc ii}] ratios correspond to two sets of theoretical ($n_e,T$) pairs mapped onto the $n_e$--$T$ plane. The values of $n_e$ and $T$ for that pixel were inferred by the intersection of the two sets of loci. Without the corrections for extinction, the median and $1\sigma$ standard deviation of the values for each jet are as follows: the electron density is $6.7\pm2.2\times10^4$ cm$^{-3}$ and a temperature of $1.7\pm0.6\times10^4$ K for the blueshifted jet, and the redshifted jet has an electron density of $4.6\pm2.4\times10^4$ cm$^{-3}$ and a temperature of $1.5\pm0.5\times10^4$ K. These values suggest that the blueshifted jet has on average $\sim46^{+59}_{-19}\%$ higher electron density and $\sim13^{+2}_{-3}\%$ higher temperature than the redshifted jet, although large scatters are present in the inferred values. The scatters in the blueshifted jet are mainly due to its low signal-to-noise, and those in the redshifted jet are mainly due to changes of physical conditions along the jet axis. From the [N {\sc ii}]/[O {\sc i}] ratio, the blueshifted jet has overall a factor of $\sim2$ -- $3$ higher ionization fraction than that of the redshifted jet. Combining the differences in electron density and ionization fraction, the hydrogen density of the redshifted jet would be a factor of $\sim1.4$ -- $2$ larger than that of the blueshifted jet. From this line ratio diagnostics, one finds that the blueshifted jet is less dense, slightly hotter and more ionized, than the redshifted jet.
        
      Uncorrected extinction possibly present in the system may introduce some more uncertainties for the inferred physical conditions. The documented $A_V$ for Sz 102 ranges from 0.32 to 2.9 \citep{Sartori03,Hughes94,MOD11,Evans09}, although what these values mean for sources seen in only scattered light is hard to assess. If $A_V=2.5$ is applied, only $3\%$ of the flux shortward of 4000\AA\ can escape from extinction, and the corrected luminosity amounts to $\gtrsim10^{30}$ erg\,s$^{-1}$ at the distance of Sz 102, much too bright for typical T Tauri stars. On the other hand, an $A_V=2.5$ is consistent with the absorption column density of the soft X-ray source close to the star (see the discussions in Section 5.3). Because the star is observed nearly edge-on, the jet likely suffers from much less extinction than does the central star. We adopt $A_V=2.5$ as the upper limit to the visual absorption. At this limit, the [N {\sc ii}] $\lambda5755/\lambda\lambda6548+6583$ would be raised by a factor of 1.4, and the [S {\sc ii}] $\lambda\lambda4068/6731$ raised by a factor of 4.5. This results in a lower temperature and much higher electron density for both jets. The median temperature and electron density of the blueshifted jet would become $9\pm2\times10^3$ K and $4\pm1\times10^5$ cm$^{-3}$, respectively, and those of the redshifted jet would be $8\pm1\times10^3$ K and $3\pm1\times10^5$ cm$^{-3}$. The asymmetry in physical conditions would appear less significant if a larger extinction is adopted. However, the data suggest that the absorption cannot be this great. At this limit we have difficulties finding an intersection in the $n_e$--$T$ plane for $\sim 30\%$ of the pixels. We adopt the inferred values without extinction corrections while noting that the derived values are strictly a lower limit for the electron density $n_e$ and a upper limit for the temperature $T$.

\section{DISCUSSION} \label{discussion}

        Sz 102 drives a pair of microjets that are peculiar compared to other microjets from low-mass YSOs. The optical [Ne {\sc iii}] $\lambda3869$ is clearly and uniquely detected in the jets. Its velocity-resolved line profile provides the unprecedented probe of the innermost region where wide-angle winds may be launched in the system. The physical conditions inferred from [Ne {\sc iii}] $\lambda3869$ and other forbidden emission lines suggest line excitation should occur in relatively hot and dense part of the flows. We discuss scenarios that would produce the neon ions and the excitation of the lines in the flows in this section. Put into perspectives, the origins of the optical [Ne {\sc iii}] emission would elucidate an important role of X-rays in the irradiation of the system. We discuss the sources of the X-rays in a system such as Sz 102 and their implications.

\subsection{Kinematics of the Sz 102 Microjets}

    The forbidden emission line profiles of the Sz 102 microjets, including the large line widths and the ``excess'' spectral features, may be explained by a wide-angle wind whose density is stratified along the jet axis, such as one in the X-wind model \citep[e.g.,][]{XWI,XWV,Shang98,Shang02,Shang10}. \citet{Shang10} illustrated synthetic line profiles of [O {\sc i}] $\lambda6300$, [Ne {\sc ii}] 12.81\micron\, and [Ne {\sc iii}] 15.55\micron\ of an approaching jet from a low-mass YSO at various inclination angles. For an approaching jet, the predicted profiles have typical full widths at zero intensity extended from $\sim -200$ to $+100$ km\,s$^{-1}$ and a blueshifted intensity peak corresponding to the projected terminal velocity of the jet. The diverging streamlines of the wide-angle wind give rise to the large line widths. The intensity peaks toward the jet terminal velocity, which comprises of largely the densest part of the flow that has collimated into jet. Part of the diverging streamlines points away from the observer near the base, allowing the radial velocities to extend to the red, producing the so-called excess emission as described in \citet{Pyo06}. As the inclination angle increases, position of the peak intensity shifts toward the systemic velocity. The shapes of the theoretical line profiles can be flipped about the systemic velocity to represent those of a receding jet. The shape and widths of the line profiles from the redshifted microjet of Sz 102 are consistent with such theoretical profiles seen close to edge-on.
        
    A wide-angle wind may account for the large blueshifted velocity centroid in Sz 102. If the approaching side of a wide-angle wind is properly ionized and excited near the base, one may be able to see through the diverging streamlines and probe the dense region that contribute to the emission. In this case, the effective viewing angle to the flow volume would be less than the inclination angle determined from radial velocities and proper motions of the outer knots after the flow has already collimated into a jet. Assume the flow speed of the blueshifted wind is similar to that determined from outer knots, $\sim320$ km\,s$^{-1}$ \citep{Krautter86}, the large blueshifted centroid would be better explained if it corresponds to streamlines directed toward the line of sight with angles $\lesssim80\degr$. The most blueshifted velocity in the profile, $\sim-300$ km\,s$^{-1}$, which can be seen from [Ne {\sc iii}] and other line profiles, is also consistent with the maximum radial velocity achievable from the blue wide-angle wind.

        Further constrained by the high critical density of [Ne {\sc iii}] $\lambda3869$, the observed wide-angle wind may come from within the innermost region of the disk of Sz 102. Consider a cylinder as an approximation to the innermost part of the jet with cylindrically stratified density profile, from within which the wind is launched and the streamlines are threaded over the outer area $A\approx 2\pi r h$. The radius of this hypothetical cylinder may be regarded as the upper limit of the cylindrical distance of the wind launching region from the outflow axis. It can be estimated, in an order-of-magnitude way, if the density within the cylinder is assumed to be the inferred jet density:
 \[ R_m \sim \frac{\dot{M}_w x_e}{2\pi h v_w m_{\rm H} n_e}
        \sim \left(\frac{\dot{M}_w}{10^{-8}M_\odot\,{\rm yr}^{-1}}\right)
             \left(\frac{x_e}{0.1}\right)
             \left(\frac{v_w}{100\,{\rm km\,s^{-1}}}\right)^{-1}
             \left(\frac{n_e}{5\times10^4\,{\rm cm^{-3}}}\right)^{-1}
             \left(\frac{h}{50\,{\rm AU}}\right)^{-1}\; {\rm AU}. \]
Assuming a cylinder with a height of $h \sim 50$ AU, mass-loss rate $\dot{M}_w \sim 2\times10^{-8}$ $M_\odot\,{\rm yr}^{-1}$, $v_w \sim 300$ km\,s$^{-1}$ the maximum wind speed, inferred electron density $n_e \approx 5\times10^4$ cm$^{-3}$, and $x_e \sim 0.3$ the ionization fraction of the wind, one derives $R_m \sim 2$ AU. The uncertainties introduced by $n_e$ from $A_V$ determination and by $x_e$ from the [N {\sc ii}]/[O {\sc i}] ratio would be of the same order of magnitude. Line profile modeling of a wide-angle wind with cylindrical stratified density suggested that most of the MIR neon emission corresponding to spectral bins that builds up the large line width originates from within cylindrical distances 1 AU from the jet axis \citep{Shang10}. 

\subsection{Physical Conditions of the Sz 102 Microjets}

        The detection of [Ne {\sc iii}] $\lambda3869$ from Sz 102 makes it possible to estimate the physical conditions for doubly ionized neon emitting gas for the first time. The IRS [Ne {\sc iii}] 15.55\micron\ flux is $2.3\times10^{-15}$ erg\,cm$^{-2}$\,s$^{-1}$ \citep{Lahuis07} and the [Ne {\sc iii}] $\lambda3869$ flux is $7.4\times10^{-15}$ erg\,cm$^{-2}$\,s$^{-1}$. At the thermal limit ($n_e \gtrsim n_{\rm cr}$), one may approximate the [Ne {\sc iii}] $\lambda3869/15.55\,\micron$ ratio as a function of $T$:
\[ \frac{L({\rm 3869\,\AA})}{L({\rm 15.55\,\micron})} = 2000 e^{-30945/T}, \]
which would be valid for $n_e \approx 10^6$--$10^7$ cm$^{-3}$. The line ratio is $\sim3$ and $\sim100$ for adopted extinctions of $A_V=0.0$ and 2.5 mag, respectively, which corresponds to temperatures of $\sim5000$ and $\sim10,000$ K. This order-of-magnitude estimate of the temperature for the neon emitting gas is consistent with that derived from [N {\sc ii}] and [S {\sc ii}] line ratios emitted from the jet. This consistency implies that the emitting volumes for [Ne {\sc iii}] and other forbidden emission may share similar physical conditions, which sustains the temperature and density that are needed for collisional excitation of the lines.

        Properties revealed by the {\sc Uves} spectra suggest that asymmetry in the Sz 102 jet persists down to scales within $\sim50$ AU. At a scale beyond $\sim 10\arcsec$ (or $\sim 2000$ AU), the redshifted outflow appears brighter and more extended than the blueshifted outflow \citep{Cof10}. Long-slit spectroscopy of the outer knots at $\sim 30\arcsec$ revealed a blue-to-red speed ratio of $\sim 2$ \citep{GH88}. This velocity asymmetry persists in the near-infrared [Fe {\sc ii}] spectra \citep{Cof10} and optical spectra from {\it HST}/STIS \citep{Cof12} and {\sc Uves}. The asymmetry in excitation conditions, inferred from the {\sc Uves} spectra, can be combined with the velocity asymmetry to elucidate physical conditions of the jets. The blueshifted jet is more ionized but less dense. It is brighter in those highly ionized lines compared with the redshifted jet, but less extended since the density drops off faster. The smaller density may in turn lead to a higher temperature if equal energy is dissipated in the same volume. 
        
        Similar asymmetries have also been observed in several other T Tauri jets \citep{Hirth94}. RW Aur A \citep{LS12} and FS Tau B \citep{Liu12}, in which 60\% and 30\% differences in velocities were found, are two well known systems driving asymmetric bipolar jets. In the context of the X-wind model, different mass loading operates on the opposite side of the disk in response to the asymmetric magnetic configuration as was suggested in the case of the RW Aur A system \citep{LS12}. The mechanism automatically adjusts the dynamics across the disk such that the averaged linear momentum is conserved in the system and the density ratio is approximately the reciprocal of the velocity ratio. The reasons for the asymmetry, however, are largely unknown, and perhaps are attributable to the detailed dynamo process inside a specific star. Other causes such as misaligned magnetic and rotation axes may also be plausible but will introduce other observable traces.

\subsection{Origins of the [Ne {\sc iii}] $\lambda3869$ Emission}

        Detection of [Ne {\sc iii}] $\lambda3869$ in the Sz 102 microjets posts strong constraints not only on physical conditions of the flow where the specific line is excited but also on the sources of ionization of neon up to the doubly ionized state. From discussion in the previous subsections, one can infer that the optical [Ne {\sc iii}] $\lambda3869$ line itself is most likely excited locally {\it in situ} inside the denser and hotter part of wide-angle winds, along with other detected forbidden transitions. How the neon is ionized and how it remains in the doubly ionized state place stringent requirements on the sources of ionization and the properties of the local flow. For the considerations of producing doubly ionized neon in an environment of low-mass YSO, one naturally looks into {\it in situ} ionization by fast shocks \citep{HRH87,HG09}, external irradiation by EUV \citep{GH08,HG09}, and X-rays \citep{GNI07,Shang10}, for sources of high-energy radiation. We discuss how each of the mechanisms may work in a system such as Sz 102.

        Neon can be collisionally ionized {\it in situ} by strong shocks that heat the flow to temperatures exceeding those of the ionization potentials. For Ne$^{2+}$, it is 62.5 eV, or $\sim 0.73$ MK. To reach a postshock temperature of $T\approx0.14[v_s/100\,\mbox{km\,s$^{-1}$}]^2$ MK \citep{Shock_ARAA93} requires a shock velocity $v_s$ of $\sim230$ km\,s$^{-1}$. Ionization from UV photons produced by the shock may reduce the required shock speed to $\gtrsim100$ km\,s$^{-1}$ \citep{HRH87,HG09}. This may account for the [Ne {\sc iii}] $\lambda3869$ detections from several large bow shocks of HH objects \citep[e.g., HH 1, 2, 32, 34;][]{BBM81,DBS82,Morse93}. The inferred shock speed of the large bow shocks in HH 34, for example, is $\sim140$ km\,s$^{-1}$ \citep{Morse92}. [Ne {\sc ii}] 12.81\micron\ detection at the northwestern knot $\sim2\farcs5$ from T Tau S suggests that shocks would be important in line formation at large distances from the star \citep{vB09}. In the case of Sz 102, the almost edge-on orientation of the system should have projected the radial component of the needed jump to be $\sim 10$ -- $20$ km\,s$^{-1}$, which may be resolved but is not present in the PV diagrams. Other signatures of strong shocks $\gtrsim100$ km\,s$^{-1}$ near the bases of the jets have not been detected or resolved up to date. It might also be that the places where these shocks occur are heavily extincted. Fast shocks would heat the gas up to 0.1 -- 1 MK, and emit in the UV to X-ray wavelengths (see below for discussions on the X-ray source detected in Sz 102). An average gas temperature of $2\times10^4$ K is inferred in this work from optical line ratios without the correction for extinction. If an extinction correction of 2.5 mag is adopted, the inferred temperature would be lowered to $9\times10^3$ K. Excitation temperature inferred from [Ne {\sc iii}] $\lambda3869$/15.55\micron\ also suggests a range between $5\times10^3$ and $10^4$ K with possible extinction corrections. The observables of the Sz 102 system thus far do not seem to support an origin of {\it in situ} ionization by fast shocks. This would imply that the ionization of neon may be done by some external high-energy photons and the local conditions of the flow will affect the excitation and formation mechanisms of the lines.

          The ionization state of the neon depends on the local dynamical conditions of the flow and how fast recombination is competing, if ionization of the neon happens away from the line-emitting volume. For the inferred temperature $T\sim2\times10^4$ K and electron density $n_e\sim6\times10^4$ cm$^{-3}$, the recombination timescale is one-half to one year. A flow speed of $\sim300$ km\,s$^{-1}$ can maintain the frozen-in neon ionization on a scale of up to 50 AU if the ionization takes place deeply inside. This is compatible with the current data set. If one adopts an extinction correction of $A_V=2.5$, the inferred lower temperature and higher density reduce the neon recombination timescale by nearly an order of magnitude. This would further disfavor strong shocks operating {\it in situ} as they would have to repeatedly ionize the neon on timescales of several months, and one would easily see tell-tale traces of the events, such as very bright knots adjacent to the driving source that resemble those at the terminal bow shocks.
To sustain the ionization of neon in the flow, temporally repetitive or spatially extended sources of ionization are in fact preferred by the evidences shown, and those would not significantly heat up the gas in the jets above a few tens of thousand degrees as inferred in this work.
        
        EUV (13.6 eV $<$ $h\nu$ $<$ 100 eV) emission from hot spots produced by accretion shocks is one source that can remove the valence electrons of neon. The EUV photons are easily absorbed by neutral hydrogen in the interstellar medium, and direct observation and determination of the ionizing flux are difficult.  For sources without strong jet activities, the existence of EUV irradiation can be obtained by MIR line diagnostics such as [Ar {\sc ii}]/[Ne {\sc ii}] and [Ne {\sc iii}]/[Ne {\sc ii}] ratios \citep{GH08,HG09}, or by free--free emission and radio hydrogen recombination lines from the disk \citep{PGH12}. Among the few MIR {\it Spitzer} [Ne {\sc iii}] disk (jet-free) sources, the generally small MIR [Ne {\sc iii}]/[Ne {\sc ii}] ratios are not positively (nor strongly) correlated with the X-ray luminosity $L_{\rm X}/L_{\rm bol}$ \citep{Esp13}. However, in jet-driving sources, the models based on the X-winds would have also predicted a small ratio of [Ne {\sc iii}]/[Ne {\sc ii}], but show correlations with X-ray parameters \citep{Shang10}. The MIR neon lines alone can not definitely establish the sources of ionization, especially in YSOs wherein both X-rays and EUV radiations may well be present. In a source where the mass accretion rate is larger than $1\times 10^{-8}M_\odot$\,yr$^{-1}$ \citep{HG09}, not only would the EUV photons have trouble escaping from the winds, but also be absorbed by the accretion funnels before they reach the inner winds \citep{GNI07,MGN08,Shang10}. In a system such as Sz 102 in which the ionization is seeded by external radiation sources, any EUV radiation that is able to penetrate the winds will simply add on top of the ionization that has been produced by the primary source of irradiation.

      Sz 102 has been known as a soft X-ray ($kT_{\rm X} < 1$ keV) emitter \citep{PJC_Gudel}. Analysis of the archival X-ray spectra of Sz 102 obtained with both {\it XMM-Newton} \citep[2003 September;][]{Xray2006} and {\it Chandra} \citep[2009 June;][]{PJC_Gudel} show soft X-ray profiles which peak at $\sim 0.6$ keV. Spectral energy fitting to the two X-ray spectra shows that the emission is produced by a plasma of $\sim 2$ MK ($kT_{\rm X} \approx 0.173$ keV) attenuated by hydrogen column of $4$--$5\times10^{21}$ cm$^{-2}$, which corresponds to $A_V \sim 2.5$ \citep[$N_{\rm H} \sim 2\times10^{21}\,A_V$ cm$^{-2}$,][]{Vuong03}. The recovered unabsorbed X-ray flux is $2.3\times10^{-13}$ erg\,s$^{-1}$\,cm$^{-2}$, or an X-ray luminosity of $1.1\times10^{30}$ erg\,s$^{-1}$ at a distance of 200 pc. This X-ray component is spatially indistinguishable from the Two Micron All Sky Survey source associated with Sz 102 within the positional uncertainty of 0\farcs4. Such a temperature, if produced in a shock, would come from an impact velocity of $\sim 380$ km\,s$^{-1}$, only achievable in the accretion funnels. The alternative would be corona trapped in the closed magnetic loops, which is known to produce soft X-ray emission in quiescence as well. \citet{CF10} noted a large equivalent width of H$\alpha$ and a maximum redshifted tail up to $+450$ km\,s$^{-1}$ in several permitted lines but absent in the forbidden lines, which they attributed to magnetospheric accretion \citep[e.g.,][]{MCH98}. A mass of 0.6--0.9 $M_\odot$ was derived, and an accretion rate of 4.2--6.3 $\times 10^{-8}$ $M_\odot$\,yr$^{-1}$ was estimated by using width of Ca {\sc ii} triplets in \citet{Com03}. If this X-ray component is produced by magnetospheric accretion, the contribution of the EUV photons would be superseded in this scenario. If the soft X-ray component is rooted in the magnetic loops, it may be easily absorbed by the accretion funnels or the bases of the winds. Regardless of the production mechanisms, the existing observations may contain contributions from both. The derived total X-ray luminosity is typical of the time-averaged value from the characteristic state \citep[defined as typical emission state between isolated flares by][]{Wolk05_COUP} of revealed low-mass pre-main sequence stars \citep[$\sim2\times10^{30}$ erg\,s$^{-1}$; see, e.g.,][]{FGP02,Wolk05_COUP}. However, a thermal spectrum at a temperature of 0.6 keV does not produce enough X-ray photons at the peak of the neon photoionization cross section \citep[$\sim0.9$ keV,][]{GNI07} for the level of the optical [Ne {\sc iii}] emission reported in this work. 

     The flat-spectrum source DG Tau, famous for its bright optical and infrared jet, tells a different story. It has a ``two-absorber'' X-ray spectrum that contains two unrelated emission components. There is a hard flaring component located at the stellar position and a soft static component elevated at $\sim0\farcs2$ along the jet axis \citep{SS08}. An extended soft X-ray ``knot'' is identified at $\approx 5\arcsec$ further down \citep{Gudel08,Gudel12}. This knot has a proper motion of $\sim0\farcs28$\,yr$^{-1}$, similar to those of the bright optical [S {\sc ii}] and UV H$_2$ knots \citep{DGTau_RadioJet,Sch13}. The X-ray knot seems to originate in the inter-knot region that has a velocity jump of $\sim100$ km\,s$^{-1}$ identified in \citet{LCD00}, if tracing back in time and position locations. The inferred temperature of the soft X-ray knot is $\sim 2.7$ MK, which is $\sim 1$ MK lower than the inner soft component \citep{Gudel12}. Based on the similar hardness, \citet{Gudel08} suggested that the inner soft component originates in the base of the outflow. \citet{GML09} showed that the temperature of the inner soft component requires strong shocks of $\sim450$ km\,s$^{-1}$ produced by the fastest portions of the jet of density $\sim 10^5$ cm$^{-3}$. The lower-temperature outer knot may be produced by collisions with previous ejecta. This scenario is consistent with its weaker shock strength and proper motion \citep{Gudel08,GML09}. The strong shocks of $\gtrsim300$ km\,s$^{-1}$ inferred along the DG Tau jet may provide some {\it in situ} ionization of neon and further excitation of the neon forbidden lines. If the luminosity of the shock-induced soft X-ray allows for enough ionization, extended neon emission and ``knot'' coinciding with the X-ray emitting jet and knot would be detected. The caveat here is that the hard flaring component is more luminous than the soft X-ray jet by almost two orders of magnitude, and even if any extended neon emission is detected along the jet, it would still not be a clean case for shock ionization. 

\subsection{X-Ray Flares in Sz 102}

     A potential source of ionizing radiation in Sz 102 is (unseen) hard X-rays. Hard ($kT_{\rm X} \gtrsim 1$ keV) X-rays can penetrate the denser region of the flows, and have sufficient ionizing power. How might such a component, if it exists, remain undetected? This component could be obscured by the optically thick disk, or it could simply be temporally variable, with the extant X-ray observations insufficient to have caught any large flares. The fact that [Ne {\sc iii}] $\lambda3869$ has been caught by two observations $\sim8$ yr apart shows that either the hard sources have flared during the two observations ({\sc Uves} and SMARTS) or the ionization has remained frozen-in within the flow and the temperature has not cooled during the time lapsed. Searching for the missing flaring components in Sz 102 may be helpful in identifying the ionization sources of its jet.

     Flares in YSOs can generate copious X-rays that are harder than those generated by plasma gyrating around the closed magnetic loops of the magnetosphere or those produced by magnetospheric accretion. Reconnection events release energies from the twisted magnetic field lines and can heat up the gas to 10 -- 100 MK \citep[as shown in, e.g., Figure 42 of][]{PF02}, whose thermal spectra are peaked at keV ranges \citep{Wolk05_COUP,Favata05_COUP}. Studies of our Sun show that flares may occur on small scales close to the star or on larger sizes extending up to the scale of the solar radius \citep[see the review in, e.g.,][]{PF02}. The former is usually called the impulsive flares, which are more frequent and less luminous, and the latter is associated with the coronal mass ejections or prominence eruptions, which release a large amount of energy. Flares in YSOs have been monitored by various X-ray satellites, and have been found to show similar characters, with elevated strengths, to those on the Sun. \citet{Wolk05_COUP} found those similar to the impulsive flares can have an average luminosity of $6\times10^{30}$ erg\,s$^{-1}$, with rise times of a day or two. Large flares are relatively rare in YSOs, but they do exist, and can have energy releases on the order up to $\sim 10^{32}$ erg\,s$^{-1}$ and temperatures exceeding 100 MK at their peak intensities \citep[see, e.g.,][]{Imanishi01,Grosso04,Favata05_COUP,Getman08a_COUP,Getman08b_COUP}. They have been interpreted to involve large loops of sub-AU sizes, which are believed to connect to both the star and the inner circumstellar disk \citep{Favata05_COUP}. This picture supports the analogies of ``coronal mass ejections'' that take place around the ``helmet streamers'' in star--disk interacting systems \citep{Shu97Sci}. These large though less frequent flares that arise high above the disk plane may release enough energy to ionize enough neon to their doubly-ionized states in the surrounding wide-angle winds (irrespective of their detailed origins). The discussions on the relative fast flow time and not-so-fast recombination time scales would help in maintaining the ionization that has been seeded by irradiation.

     A flaring hard component, similar to that detected at the stellar position of DG Tau, may help. {\it Chandra} observations during 2004 and 2006 showed that this component has a temperature ranging from 25 to 35 MK, and {\it XMM-Newton} observations in 2004 showed an even higher average temperature of $\sim70$ MK \citep{Gudel08}. Temporal analysis revealed 20 ks scale flares in the hard component \citep{Gudel07} that can recur in days \citep{Gudel12}. During the rise of the flare, energy can be released at beyond a rate of $L_{\rm X}\approx10^{30}$ erg\,s$^{-1}$; the gas is heated to $\gtrsim 100$ MK and the X-ray spectrum is hardened between 1 and 10 keV \citep{Gudel07}. This deeply embedded ($N_{\rm H}\approx2\times10^{22}$ cm$^{-2}$) X-ray source is interpreted as hot coronal gas confined in closed loops of magnetosphere that is heated by recurring flares \citep{Gudel07,Gudel08}. For Sz 102, an analog of this recurring hard X-ray component would be an ideal source of irradiation. The flux-calibrated SMARTS spectrum observed in 2011 has a [Ne {\sc iii}]$\lambda3869$ flux of $4.9\pm1.0\times10^{-15}$ erg\,s$^{-1}$\,cm$^{-2}$. This corresponds to a $\sim35\%$ drop of flux compared to the {\sc Uves} spectrum observed in 2003. If no fresh ionization has been supplied during the two events, this implies a recombination timescale of $\sim20$ yr for the doubly-ionized neon. This may be achieved if the gas in flows stays at a temperature of $\sim10^4$ K and the electron density is below $2\times10^3$ cm$^{-3}$. However this density would be too low for the observed [Ne {\sc iii}] $\lambda3869$ and other high-density lines such as [S {\sc ii}] $\lambda4068$. Although no clear history of the optical [Ne {\sc iii}] flux variation can be made from existing data, at least another event of strong flare prior to the 2011 SMARTS observations would be needed.

      Whether the occurrence and energetics of the flares in the system are asymmetric on the opposite sides of the disk remains unclear. As discussed in Section 5.2, different mass loading in the magnetocentrifugal winds may further introduce different velocity and density profiles across the disk \citep[e.g.,][]{LS12}, which subsequently will affect the ionization fractions, if the flares are symmetric (i.e., same luminosity). If the asymmetry exists in the magnetic field configurations across the disk plane, the energetics of the flares may also be asymmetric. However, the asymmetric character of the system is not required for the scenarios of external irradiation; whether it can enhance the level of X-ray flares will be investigated in the future.

\section{SUMMARY} \label{summary}

     The velocity-resolved [Ne {\sc iii}] $\lambda3869$ emission is detected and reported for the first time in the bipolar microjets of Sz 102. The optical [Ne {\sc iii}] emission traces the blueshifted and redshifted microjets that are spatially separated by a projected distance of $\sim 60$ AU. This is also the first [Ne {\sc iii}] $\lambda3869$ detection close to a low-mass jet-driving YSO. [Ne {\sc iii}] $\lambda3869$ has the largest line width and highest critical density among detected optical forbidden lines. The large line width extending to $\sim-300$ km\,s$^{-1}$ and the extended emission wing across systemic velocity can be explained by the diverging streamlines toward the observer, characteristic of a wide-angle wind \citep{Shang10}.

     Kinematics and physical conditions of the Sz 102 microjets inferred from properties of the optical [Ne {\sc iii}] line and other forbidden lines suggest that the system is driving a pair of asymmetric jets. The velocity centroids of the blueshifted jet is larger than those of the redshifted jet by a factor of 2 -- 4, and the overall line widths in the blueshifted jet is larger. The redshifted jet is overall brighter than the blueshifted jet. Optical emission-line diagnostics suggest the jets have an average temperature $T\lesssim2\times10^4$ K and electron density $n_e\gtrsim6\times10^4$ cm$^{-3}$. The contrast of line intensities are consistent with blueshifted jet being slightly hotter and more ionized. Such asymmetry in the jet has been detected in several T Tauri jets, such as FS Tau B \citep{Liu12} and RW Aur A \citep{LS12}. 
         
     The bright optical [Ne {\sc iii}] forbidden line puts a stringent constraint on how neon is ionized in the Sz 102 jets. Ionization mechanisms, such as {\it in situ} shock ionization beyond $\sim100$ km\,s$^{-1}$, external irradiation by EUV and hard X-rays, are discussed in the context of the Sz 102 system. The inferred temperature from optical line ratios ($\sim 10^4$ K) is incompatible with that of strong shocks ($\gtrsim 10^6$ K), suggesting the ionization of neon and excitation of the optical [Ne {\sc iii}] line may not occur in the same volume of the jet. Soft X-rays as well as EUV radiations produced in accretion shocks may be easily absorbed and would not provide enough hard photons for the doubly ionized neons. Large but rare flares are known to generate ample keV photons in other YSO systems such as DG Tau. Although undetected in Sz 102 thus far, they may meet the energetic requirement of the ionizing photons. Future X-ray monitoring of the Sz 102 system is key to confirm the existence of the hard X-ray component. Further higher angular-resolution optical observations are also needed to trace the ionization and heating history of the microjets. To accomplish this mission, combined observations with {\it Chandra} and {\it HST} are ideal.

\acknowledgments
        The authors would like to thank the anonymous referee whose constructive comments help improve the clarity and strengthen the conclusions of the manuscript. This work was supported by funds from the Academia Sinica Institute of Astronomy and Astrophysics (ASIAA), and the National Science Council (NSC) of Taiwan by grants NSC 100-2917-I-002-050, NSC 100-2112-M-001-004-MY2, and NSC 102-2119-M-001-008-MY3. This work is partly based on observations made with ESO Telescopes at the La Silla Paranal Observatory under programme ID 71.C-0429(C). Some of the spectra discussed herein were obtained on the CTIO 1.5m telescope, operated by the SMARTS Consortium. F.M.W. is able to participate in the SMARTS Consortium through a generous grant from the Provost of Stony Brook University.

{\it Facilities:} \facility{CTIO:1.5m (RCSpec)}, \facility{VLT:Kueyen ({\sc Uves})}

%%%%%%%%%
\clearpage

\begin{deluxetable}{llrrc} 
\tablecolumns{5} 
\tablewidth{0pc} 
\tablecaption{Low-Dispersion Line Intensities of Sz 102 Relative to H$\beta=100$\label{tbl-blow}}
\tablehead{ 
\colhead{$\lambda_{\rm air}$} &
\colhead{Line ID} & \colhead{This Work} & \colhead{\citet{Krautter84}} & \colhead{Remarks} \\
\colhead{(\AA)} & \colhead{} & \colhead{($R\sim1000$)} & \colhead{($R\sim500$)} & \colhead{}}  
\startdata 
3727     & [O {\sc ii}]     &   72 & \nodata & \\
3797.900 & H10              &    5 & \nodata & \\
3835.000 & Mg {\sc i} (3)   &    9 & \nodata & \\
3868.752 & [Ne {\sc iii}]   &   12 & \nodata & \\
3889.050 & H8               &   14 & \nodata & \\
3933.673 & Ca {\sc ii} K    &   42 & 139     & \\
3968.487 & Ca {\sc ii} H    &   40 & 138     & 
           blended with [Ne {\sc iii}] $\lambda3968$ and  H$\epsilon$ \\
4072.000 & [S {\sc ii}]     &   91 & 135     & \\
4101.740 & H$\delta$        &   20 &  20     & \\
4244.4   & [Fe {\sc ii}]    &    8 &  21     & \\
4340.470 & H$\gamma$        &   40 &  58     & \\
4363.000 & [O {\sc iii}]    &   13 & \nodata & \\
4414.500 & [Fe {\sc ii}]    &   10 &  21     & \\
4472.929 & Fe {\sc ii}      &    5 & \nodata & \\
4571.000 & Mg {\sc i}]      &    2 &  16     & \\
4658.3   & [Fe {\sc ii}]    &    4 &   8:    & \\
4731.453 & Fe {\sc ii}      &    2 & \nodata & \\
4814.600 & [Fe {\sc ii}]    &    5 &   9:    & \\
4861.330 & H$\beta$         &  100 & 100     & \\
4893.820 & Fe {\sc ii}      &    2 & \nodata & \\
4923.927 & Fe {\sc ii}      &    2 & \nodata & \\
4959.000 & [O {\sc iii}]    &   12 &   7:    & \\
5006.800 & [O {\sc iii}]    &   37 &  24     & \\
5158.4   & [Fe {\sc ii}]    &   17 &  11     & \\
5200.700 & [N {\sc i}]      &    7 &   3:    & \\
5261.600 & [Fe {\sc ii}]    &    9 &   9:    & \\
5273.400 & [Fe {\sc ii}]    &    8 & \nodata & 
\enddata
\end{deluxetable}

\begin{deluxetable}{clrccccc} 
\tablecolumns{8} 
\tablewidth{0pc} 
\tabletypesize{\footnotesize}
\tablecaption{Properties of Selected Forbidden Emission Lines from Sz 102\label{LineProperties}}
\tablehead{ 
\colhead{Panel} &
\colhead{Line} & \colhead{I.P.} & \colhead{$E_u$} & \colhead{$n_{\rm cr}$ ($10^4$ K)} & 
\colhead{PSF} & \colhead{$\sigma$ level} & \colhead{Contours} \\
\colhead{} & \colhead{} & \colhead{(eV)} & \colhead{(K)} & \colhead{(cm$^{-3}$)} & 
\colhead{(\arcsec)} & \colhead{(erg\,cm$^{-2}$\,s$^{-1}$\,\AA$^{-1}$)} & \colhead{($\sigma$)}}
\startdata 
(a) & [Ne {\sc iii}] $\lambda3869$ & 62.526 & $3.7\times10^4$ & $1.1\times10^7$ & 
                     $1.71$ & $5.1\times10^{-18}$ & 10, 20, 30, 40, 50 \\
(b) & [N {\sc ii}]   $\lambda6583$ & 14.534 & $2.2\times10^4$ & $8.1\times10^4$ &
                     $1.33$ & $2.8\times10^{-18}$ & 10, 20, 30, 50, 75, 100, 150, 200, 300 \\
(c) & [O {\sc iii}]  $\lambda5007$ & 48.735 & $2.9\times10^4$ & $5.7\times10^5$ &
                     $1.42$ & $4.3\times10^{-18}$ & 10, 20, 30, 40, 50, 75, 100 \\
(d) & [O {\sc ii}]   $\lambda3726$ & 13.618 & $3.9\times10^4$ & $3.8\times10^3$ &
                     $1.71$ & $1.4\times10^{-17}$ & 10, 20, 30, 40, 50 \\
(e) & [O {\sc i}]    $\lambda6300$ &  0.000 & $2.3\times10^4$ & $1.4\times10^6$ & 
                     $1.27$ & $2.7\times10^{-18}$ & 10, 20, 30, 50, 75, 100, 200, 500, 1000 \\
(f) & [S {\sc ii}]   $\lambda4068$ & 10.360 & $3.5\times10^4$ & $2.0\times10^6$ &
                     $1.70$ & $1.6\times10^{-17}$ & 10, 20, 30, 50, 75, 100, 150 \\
(g) & [S {\sc ii}]   $\lambda6731$ & 10.360 & $2.1\times10^4$ & $2.7\times10^3$ &
                     $1.33$ & $9.6\times10^{-19}$ & 10, 20, 30, 50, 75, 100, 200, 500, 1000 
\enddata
\tablecomments{
 \begin{itemize}
 \itemsep=-1pt
 \item I.P.: ionization potential required from its neutral species. 
 \item $E_u$: upper-level energy relative to the ground state of the configuration of that species. 
 \item $n_{\rm cr}$: critical density at $T = 10^4$ K.
 \item PSF: Gaussian FWHM of the point spread function.
 \end{itemize}
}
\end{deluxetable}

\end{document}